\documentclass[12pt]{iopart}
\pdfoutput=1
\usepackage[latin9]{inputenc}
\usepackage{textcomp}
\usepackage{graphicx}
\usepackage{esint} 
\usepackage{color}
\usepackage{cite}

\usepackage{color}
\usepackage{soul}

\makeatletter

\providecommand{\tabularnewline}{\\}

\usepackage{iopams}
\usepackage{setstack}

\makeatother

\begin{document}

\title{Frontier molecular orbitals of single molecules adsorbed on 
thin insulating films supported by a metal substrate: A simplified
density functional theory approach}

\author{Iv\'{a}n Scivetti and Mats Persson}

\address{Surface Science Research Centre and Department of Chemistry
University of Liverpool Liverpool L69 3BX UK}

\ead{mpersson@liverpool.ac.uk}

\maketitle
\date{\today} 
\begin{abstract}
We present a simplified density functional theory (DFT) method to
compute vertical electron and hole attachment energies to frontier
orbitals of molecules absorbed on insulating films supported by a
metal substrate. The adsorbate and the film is treated fully within
DFT, whereas the metal is treated implicitly by a perfect conductor
model. As illustrated for a pentacene molecule adsorbed on NaCl films
supported by a Cu substrate, we find that the computed energy gap
between the highest and lowest occupied molecular orbitals - HOMO and
LUMO -from the vertical attachment energies increases with the
thickness of the insulating film, in agreement with experiments. This
increase of the gap can be rationalized in a simple dielectric model
with parameters determined from DFT calculations and is found to be
dominated by the image interaction with the metal. However, this model
overestimates the downward shift of the energy gap in the limit of an
infinitely thick film.  This work provides a new and efficient
strategy to extend the use of density functional theory to the study
of charging and discharging of large molecular absorbates on
insulating films supported by a metal substrate.  \\ \\
\\
\smallskip{}
 \medskip{}

\end{abstract}



\section{Introduction}\label{sec:intro} 
A remarkable capability of scanning probe microscopy is the
possibility to form and control different (meta-)stable charged states
of single atoms and molecules adsorbed on thin insulating films
supported by a metal substrate
\cite{repp1,olsson,repp2,repp3,mohn1,wolfram1,bruno1,pav,mohn2,moll,gross1,majzik,wolfram2,wolfram3,bruno2,gross2,
  bruno3}.  These charged states can be manipulated and characterized
on the atomic scale by scanning tunnelling microscopy (STM)
\cite{repp1,olsson,repp2,repp3,mohn1,wolfram1,bruno1,pav,mohn2} atomic
force microscopy (AFM)
\cite{bruno1,pav,mohn2,moll,gross1,majzik,wolfram2,wolfram3,bruno2,gross2}
and Kelvin probe force microscopy (KPFM) \cite{gross2,
  bruno3}. Charged states in molecular adsorbates are formed either by
electron attachment to the lowest unoccupied molecular orbital (LUMO)
or hole attachment to the highest occupied molecular orbital (HOMO)
\cite{repp2}. Control over the occupancy of frontier molecular states
is crucial to alter selectively the catalytic properties of single
adsorbates. Such a capability plays a decisive role in the field of
molecular electronics, for example, where the ultimate goal is to use
single molecules as functional building blocks for switches,
rectifiers, transistors and memory
\cite{repp2,repp3,mohn1,wolfram3,lilj,quek,diez,yee,joachim,lortscher,ratner,schull}

Charge states of adsorbates can be stabilised either by having a
sufficently large polarization of the ionic film and the metal
substrate or a sufficiently long lifetime using a sufficiently thick
film to increase their life time. For example, the Au anion adsorbed
on a NaCl bilayer is stabilized by the large polarization of the ionic
film and the metal substrate~\cite{repp1} , whereas in the case of a
pentacene molecule adsorbed on this bilayer, this
polarization is neither sufficently large to stabilize the anion or
the cation\cite{repp2}. In contrast, for pentacene molecule on NaCl
films thicker than 13 monolayers (ML), the electron transfer through
the film is quenched and the life time is sufficiently large so that
both the cation and the anion become stable on the time scale of the
experiments \cite{wolfram3}. Nevertheless, the energies of frontier
molecular orbitals is found to depend on film thickness. In fact,
scanning tunneling spectroscopy of ultra-thin films ($\le$ 3ML) and
AFM experiments of thick films ($>$ 20ML) showed that the HOMO-LUMO
gap of adsorbed pentacene increases with increasing the film
thickness, but this gap was always smaller than its value in gas
phase. This behaviour calls for calculations to gain a deeper insight
into the interplay between metal substrate and film thickness in
determining the energies of the frontier orbitals of the molecular
adsorbates.

In principle, the electron and hole attachment energies which determine,
for instance, the observed HOMO-LUMO gap of molecular adsorbates cannot
simply be computed using ground state density functional theory (DFT)
\cite{HK,KS}, and an excited-state theory such as the
GW approximation is required~\cite{GW}. However, GW calculations are very
expensive for typical molecular adsorbates \cite{sementa} and have so
far only been applied to a CO molecule adsorbed on NaCl films
supported by a semi-conductor \cite{FreRinSche09}.

For adsorbates on thin insulating films supported by a metal substrate
even the computation of ground state energies for stable charged
states can be very challenging at the DFT level
\cite{repp1,olsson,mohn1} due to the charge-delocalisation error
introduced by current exchange-correlation functionals
\cite{DFTdisc,DFTdisc2}. This error often leads to fractional
charging, a problem which sometimes can be eliminated using a DFT+U
approach \cite{anisimov,cococcioni}. This approach has been
successfully applied to the calculation of multiply charged states of
Ag adatoms~\cite{olsson}.

A simplifying feature of this class of system is the insulating
character of the ionic film, which significantly reduces the coupling
of the adsorbate electronic states with the metal electronic states,
and forms the basis for approximate schemes
\cite{aupenta_TB}. This weak coupling was considered previously in DFT
calculations of electron attachment energies to vacancy states in a NaCl
bilayer on a Cu surface by constraining the vacancy state to be
occupied \cite{gross2}. Recently, we developed an approximate method
where the metal electrons are eliminated completely but the ionic film
and the adsorbate are treated fully within DFT
\cite{ISMP_1,ISMP_2}. The metal is simply replaced by a perfect
conductor (PC) model and the remaining non-Hartree interactions
between the metal substrate and the film are modelled by a simple
force field (FF) whose parameters are obtained from full DFT
calculations of the ionic film supported by the metal substrate. By
construction, this new method (DFT-PC-FF) makes it possible to control
the charge state of the adsorbate with a large reduction of the
computational time. This method was applied successfully to the
calculation of the Au anion on a NaCl bilayer supported by a Cu
substrate \cite{ISMP_2}.

In this work, we show how DFT-PC-FF simulations can be used to compute
the energies of frontier molecular orbitals of adsorbates on
insulating films supported by a metal substrate. As an example, we
consider pentacene adsorbed on NaCl films of different thicknesses
which are supported by a Cu substrate.  In agreement with experiments,
we find that the HOMO-LUMO gap increases with increasing number of
NaCl layers. In addition, our findings are compared with the results
from a simple dielectric model of the adsorbed film with abrupt
interfaces. We show that this simplified model is able to
semi-quantitatively describe the variation of the HOMO-LUMO gap with
film thickness but has also some shortcomings.

The paper is organised as follows. In Section \ref{sec:DFT-PC} we
describe the main ingredients of the DFT-PC-FF method and its
application to the calculation of electron and hole attachment
energies to an adsorbed molecule and the corresponding HOMO-LUMO
gap. Computational details are presented in
Sec. \ref{sec:comp_methods}. Results for the electronic structure and
the HOMO-LUMO energy gap for the isolated molecule are presented in
Section \ref{subsec:isomol}. Results for the variation of this energy
gap of an adsorbed pentacene molecule on NaCl films with the number of
layers are presented and compared with experiments in Section
\ref{sec:geo-ene}. In this latter section we also compare the
calculated and experimnental energy gaps with the results from a
simple dielectric model in order to elucidate the contributions from
the film and the metal substrate. Finally, some conclusing remarks of
this work are presented in Section \ref{sec:conclusions}. Electrostatic units
are used throughout in this paper.

\section{Computation of charged states of adsorbates with DFT-PC-FF}
\label{sec:DFT-PC} 

The DFT-PC-FF method for the calculations of charged 
adsorbates on an ionic insulating film supported by a metal substrate
has been described in detail in our previous
work \cite{ISMP_1,ISMP_2}. Here, we just summarise the key points
of this method and how it can be used to calculate the
electron and hole attachment energies to an adsorbed molecule and the
corresponding HOMO-LUMO gap. 

In the DFT-PC-FF method, the electrons of the metal substrate
are not explicitly included in the calculation but their screening
is accounted for in a perfect conductor (PC) model. The residual non-Hartree
interactions between the film and metal substrate are described by
a force field (FF). The system is represented in a supercell with a prescribed
total charge $Q_{s}$ of the adsorbate and film, which is compensated by
an induced charge $-Q_{s}$ in the PC, so that the supercell is neutral. 
The total energy $E_{PC-FF}^{(Q_{s}/e)}$ of the system is then given in this method
by minimising the following energy functional 
\begin{equation}
E_\mathrm{PC-FF}^{(Q_s/e)}[n_{s}]=\bar{E}_\mathrm{PC}[n_{s}]- \frac{Q_{s}}{e}\Phi_\mathrm{PC}+\sum_{k\in\mathrm{FL}}\phi_{k}(z_{k})\label{eq:E_PC-FF-Def}
\end{equation}
with respect to the electron density $n_{s}$ of the adsorbate and the
film under the constraint that the total charge of the electrons and
the ions is $Q_{s}$ . Here, $\bar{E}_\mathrm{PC}[n_{s}]$ is the
energy functional of the adsorbate and the film system interacting
with the PC and $\Phi_\mathrm{PC}$ is the effective work function. As
detailed in Ref. \cite{ISMP_2}, the difference between
$\Phi_\mathrm{PC}$ and the workfunction $\Phi$ of the film supported
by the explicit metal substrate (with no absorbate) is due to the
overlap of the electron density of the film with the image plane in
the PC model. This overlap gives rise to a potential difference
between the PC plane and the vacuum level. The residual non-Hartree
interactions between the film and metal substrate are described by a
simple force-field (FF) based on non-polarisable pair potentials, $\phi_{k}(z_{k})$ which only
depends on the perpendicular distance $z_{k}$ between the ions $k$ in
the first layer of the film and the image plane. Besides the atom
kinds of the adsorbate and the films, the material specific parameters
in the DFT-PC-FF method are the position $z_\mathrm{im}$ of the image
plane in the PC model with respect to the metal surface plane, the
effective work function $\Phi_\mathrm{PC}$ and the pair potentials in
the FF. How these parameters were determined is described in Section
\ref{sec:comp_methods}.

In scanning tunneling spectroscopy and AFM experiments, the bias
voltages corresponding to attaching a tunneling electron or hole to
the adsorbate are determined by the transition energies $\Delta
E_\mathrm{e}$ and $\Delta E_\mathrm{h}$ for attaching an electron or a
hole from the Fermi level to the adsorbate at fixed ion-core
positions, respectively.  In the DFT-PC-FF method these energies are
approximated by the following vertical transition energies
\begin{eqnarray}
\Delta E_\mathrm{e} & = &
E_\mathrm{PC-FF}^{(Q_s^0/e-1)}-E_\mathrm{PC-FF}^{(Q_s^0/e)}\label{eq:DE_e_Def}\\ \Delta
E_\mathrm{h} & = &
E_\mathrm{PC-FF}^{(Q_s^0/e+1)}-E_\mathrm{PC-FF}^{(Q_s^0/e)}\label{eq:DE_h_Def}
,
\end{eqnarray}
where the energies of the negatively charged state $Q_s^0 - e$ and
positively charged state $Q_s^0 +e)$ are obtained at the calculated
equilibrium geometry for the molecule in its ground state with charge
$Q_s^0$ . Note that in the presence of a metal support these charge
states are not electronic ground states but are resonances due to mixing with
metallic states. However, this mixing is already very weak, for
instance, for a pentacene molecule adsorbed on a NaCl bilayer due to
its insulating character, resulting in very narrow resonances with an
estimated broadening of a few hundreds of an $\mu$eV
\cite{repp2}. Thus, the ground state energies obtained by neglecting
this mixing should be an excellent approximation to these resonance
energies. Since the system is represented in a supercell, the computed
results are expected to depend on the surface area or coverage due to
electrostatic adsorbate-adsorbate interactions. Thus, it is necessary
to correct $\Delta E_\mathrm{e}$ and $\Delta E_\mathrm{h}$ for the
dipole-dipole interactions between periodic images, as described in
\ref{app:dipcorr}.

If the adsorbed molecule is neutral ($Q_s^0=0$) as for the adsorbed
pentace molecule then the HOMO-LUMO gap can be obtained from the vertical
transition energies $\Delta E_\mathrm{e}$ and $\Delta E_\mathrm{h}$ as
follows
\begin{equation}
E_\mathrm{G}=\Delta E_\mathrm{e}+\Delta
E_\mathrm{h}=E_\mathrm{PC-FF}^{(-1)}+E_\mathrm{PC-FF}^{(+1)}-2E_\mathrm{PC-FF}^{(0)}.\label{eq:egap_Def}
\end{equation}
Note that this result for $E_\mathrm{G}$ is independent of the
(effective) work function $\Phi_\mathrm{PC}$ of the metal and the
film. For a reference, the corresponding HOMO-LUMO gap 
for an isolated neutral molecule, $E_\mathrm{G0}$,  was also calculated
from the vertical affinity $A=E^{(0)}-E^{(-1)}$ and ionisation
energies $I=E^{(+1)}-E^{(0)}$ as
\begin{equation}
E_\mathrm{G0}=I-A=E^{(+1)}+E^{(-1)}-2E^{(0)}\label{eq:egap0_Def}
\end{equation}
where $E^{(Q/e)}$ is the total energy of the isolated molecule with a
net charge of $Q$ in the equilibrium geometry of the neutral
molecule.

\section{Computational details}
\label{sec:comp_methods} 
Periodic DFT calculations were performed using the VASP code
\cite{vasp1,vasp2}.  All the required modifications for the
implementation of the DFT-PC-FF method in VASP have already been
detailed in Refs. \cite{ISMP_1} and \cite{ISMP_2}. The projector
augmented wave method (PAW) \cite{paw,vasp3} was used to describe the
electron-ion interaction with a plane wave cut-off energy of 400
eV. The electronic exchange and correlation effects were treated using
the optB86b-vdW version of the van der Waals (vdW) density functional
\cite{klimes1,klimes2}.  The NaCl bilayer supported by a Cu(100)
substrate was modelled using a slab in a supercell. As detailed in
Ref. \cite{ISMP_2}, each primitive surface unit cell was composed of
four layers of Cu atoms with nine Cu atoms in each layer, and a NaCl
film that contained four atoms of each species in each layer.  Note
that in the DFT-PC-FF simulations this unit cell only contained the
eight atoms of the NaCl film.

The Cu substrate was included explicitly just to compute the pentacene
molecule adsorbed on a NaCl bilayer in order to make a comparison with
the DFT-PC-FF results. We shall refer to this calculation as
DFT-full. In this calculation, the supercell included $3\times2$
repetitions of the primitive surface unit cell.

In the DFT-PC-FF method, the dependence on the lateral
size of the supercell was investigated for the NaCl bilayer
using supercells containing $3\times2$, $3\times3$, $4\times3$
and $4\times4$ repetitions of the primitive surface unit cell. In the
case of NaCl films thicker than the bilayer only supercells containing
the $4\times3$ surface unit cells were considered. The
lateral sizes of the supercells were sufficiently large to limit the
sampling of the surface Brillouin zone to the $\Gamma$-point.  All
ionic relaxations were carried out with a convergence criteria of 0.02
meV/\AA{} for the magnitude of the forces. The density of states 
of partial waves (PDOS) around the various atom sites were obtained using the
PAW method.

As reported in Ref. \cite{ISMP_2} for the Cu(100) substrate, the value
of 1.48 {\AA} for $z_\mathrm{im}$ was obtained from the calculated
response of this substrate to an external electric field. A value of
3.24 eV for $\Phi_\mathrm{PC}$ was obtained from the calculated
electrostatic potential and $\Phi=3.74$ eV was the computed work
function of the supported NaCl bilayer including the Cu(100) substrate
in the DFT-full simulation.  Furthermore, the functional form and the
parameters of the pair potentials are the same as for the
non-polarisable FF developed in Ref. \cite{ISMP_2}, which were
obtained from DFT calculations of the NaCl bilayer on the Cu(100)
substrate.

In order to calculate $E_\mathrm{G0}$ from
Eq. (\ref{eq:egap0_Def}) of the isolated pentacene moleculecule,
spin-polarized calculations using the Makov-Payne scheme\cite{makov}
were performed for a positively and negatively charged molecule in a
cubic supercell. Total energies including dipole corrections were
converged for a supercell with a side length of 30 \AA.

\section{Results}
\label{sec:results} 

In this section, we present the results of the DFT-PC-FF simulations
for the vertical transition energies $\Delta E_\mathrm{e}$ and $\Delta
E_\mathrm{h}$ and the corresponding HOMO-LUMO gap, $E_\mathrm{G}$, for
a single pentacene molecule adsorbed on NaCl films with different
number of atomic layers. Computed energies are compared with
experimental results. In addition, we present and discuss results for
a simplified electrostatic model of this system. We begin by
addressing the problem of an isolated pentacene molecule in vacuum.

\subsection{Isolated pentacene molecule}
\label{subsec:isomol}
\begin{figure}
\centering \includegraphics[scale=0.4]{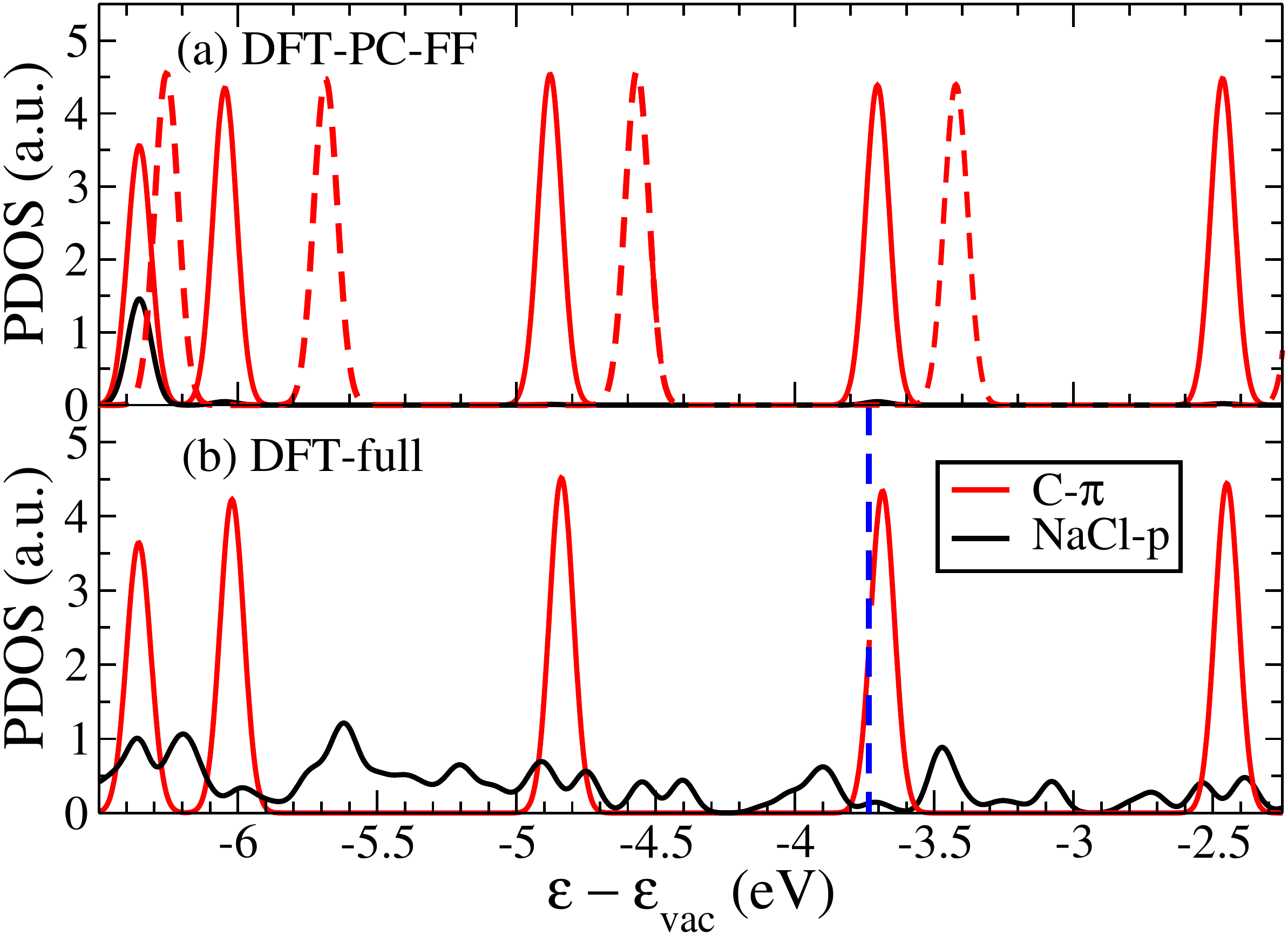}
\caption{(online color) Calculated PDOS of $p$ partial
  waves around the Na and Cl ions (solid black lines) and $p_z$ partial
  waves around all C atoms of an isolated pentacene molecule (dashed
  red lines) and a pentacene molecule adsorbed on a NaCl
  bilayer supported by a Cu(100) substrate (red lines) using (a) DFT-PC-FF and (b)
  DFT-full. All energies are referenced with respect to the vacuum level
  $\epsilon_\mathrm{vac}$. The Fermi energy in (b) is indicated by the
  vertical dashed blue line. The PDOS was broadened by a Gaussian with a
  broadening (FWHM) of 0.1 eV.}
\label{fig:pdos_penta} 
\end{figure}

As a reference case, the electronic states and the HOMO-LUMO gap
$E_\mathrm{G0}$ of the isolated pentacene molecule were also
computed. The energies of the $\pi$ orbitals of the molecule were
revealed by the calculated PDOS of $p_{z}$ states of all C atoms of
the neutral molecule, as shown by the dashed red lines in
Fig. \ref{fig:pdos_penta}(a). The calculated orbital densities of the
frontier orbitals -- HOMO and LUMO -- which are of $\pi$ character are
shown in the upper panel of Fig. \ref{fig:IR_penta}. As expected from
DFT calculations using a GGA-like approximation like the optB86b-vdW for
the exchange-correlation functional, the calculated band gap from the
computed Kohn-Sham (KS) energies of the HOMO and LUMO is 1.14 eV,
which severely underestimates the experimental value of 5.27 eV
\cite{sato,repp2}. In contrast, DFT calculations of the total energies
of the positively and negatively charged, as well as the neutral
molecule in the equilibrium geometry of the neutral molecule, as
detailed in Sec.\ref{sec:comp_methods} and using
Eq.~(\ref{eq:egap0_Def}), gave a much improved value of
$E_\mathrm{G0}$= 4.62 eV. This result is in good agreement with
the value of 4.73 eV from previous DFT calculations using the PBE
functional\cite{endres}, and the result of 4.72 eV from time-dependent
DFT calculations using the B3LYP functional \cite{tddft}. 

\subsection{Pentacene adsorbed on the NaCl bilayer supported by a Cu(100) surface}
\label{sec:geo-ene}
DFT-full calculations show that the adsorbed molecule on the NaCl
bilayer supported by the Cu(100) surface is neutral and preserves
the planar geometry \cite{quasi_planar} of the isolated free molecule 
with negligible changes in the
interatomic distances. In the most stable configuration of the adsorbed
molecule, the central aromatic ring is on top of a Cl anion with a
distance of 3.05 \AA{} from the outermost NaCl layer. This large distance
and the rather weak adsorption energy of 1.65 eV is due to the closed-shell
electronic structure of pentacene and the formation of a physisorption
bond. Results obtained using the DFT-PC-FF calculations give a
very similar molecule-surface distance of 3.06 \AA{} and an adsorption
energy of 1.68 eV. This good agreement provides strong support for
our proposed DFT-PC-FF method, especially considering also  the
massive reduction of the computational time by a factor of about 70
compared to the DFT-full calculations.

\begin{figure}
\centering %
\begin{tabular}{|c|c|}
\hline 
\multicolumn{2}{|c|}{\textbf{Gas phase}}\tabularnewline
\hline 
\textbf{HOMO}  & \textbf{LUMO} \tabularnewline
\includegraphics[scale=0.08]{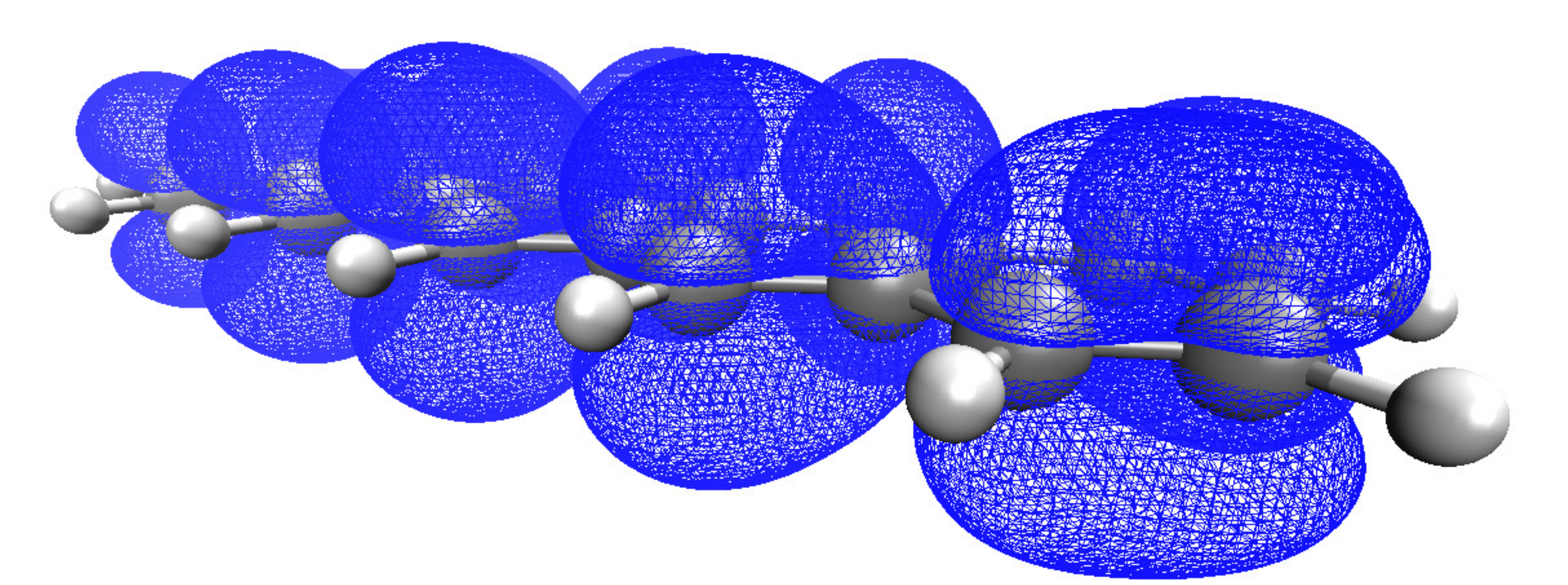}
\includegraphics[scale=0.08]{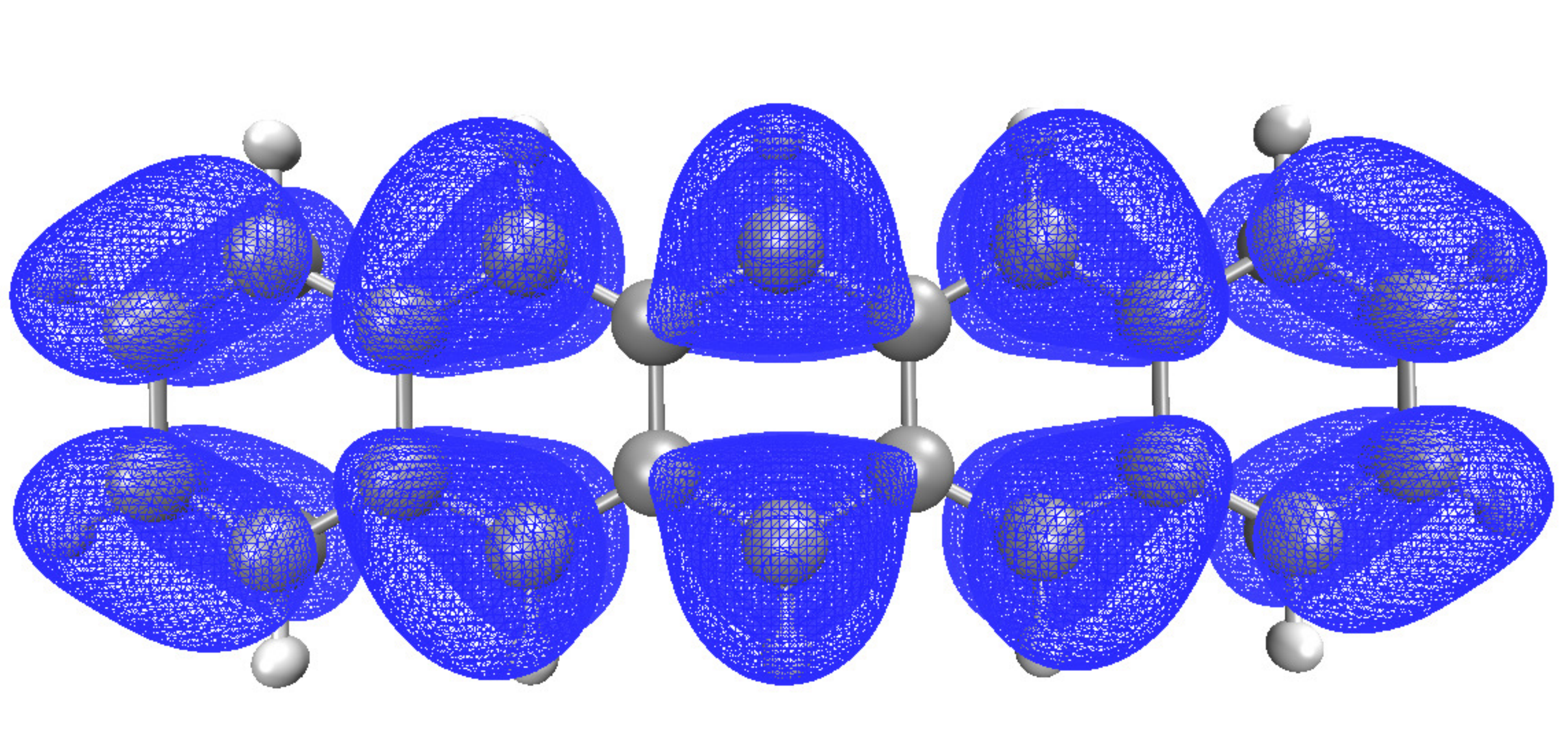}
&
 \includegraphics[scale=0.08]{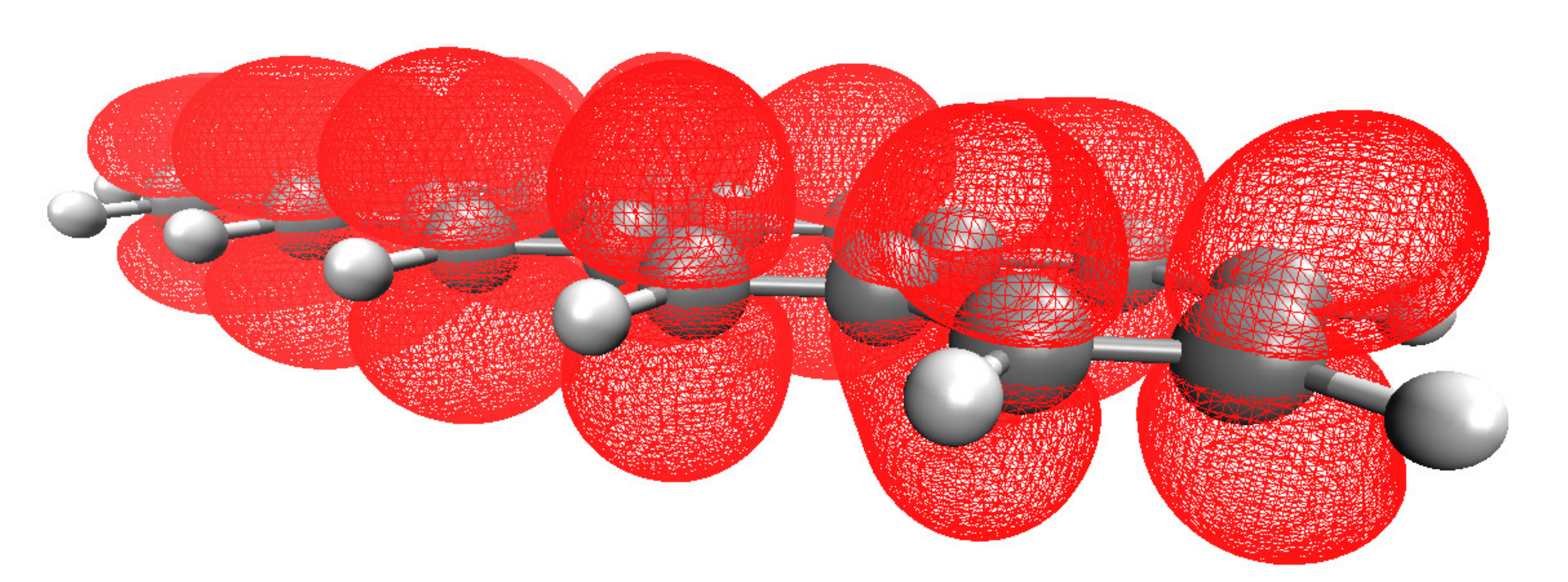}
 \includegraphics[scale=0.08]{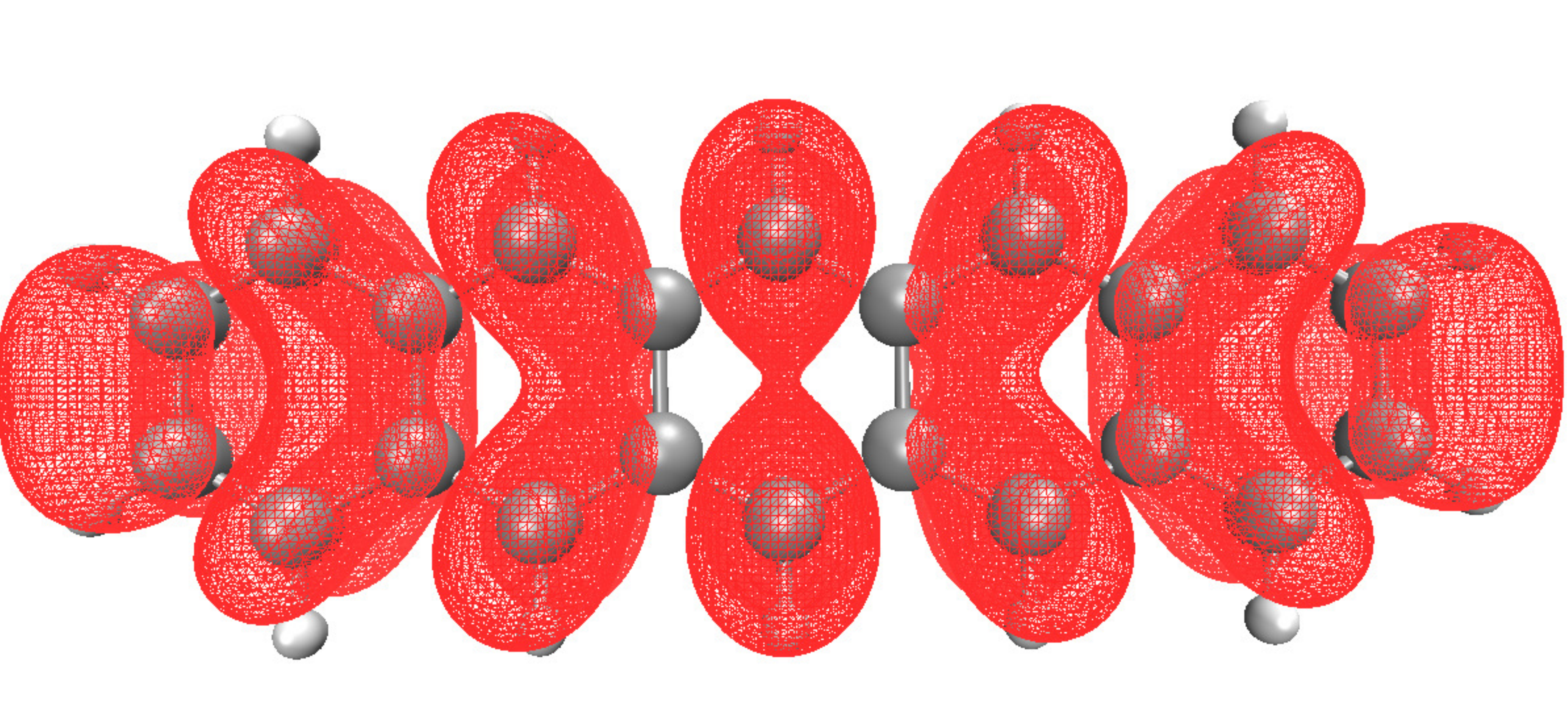}\tabularnewline
 \hline 
 \multicolumn{2}{|c|}{\textbf{Adsorbate}}\tabularnewline
 \hline 
 \textbf{HOMO}  & \textbf{LUMO} \tabularnewline
 \includegraphics[scale=0.13]{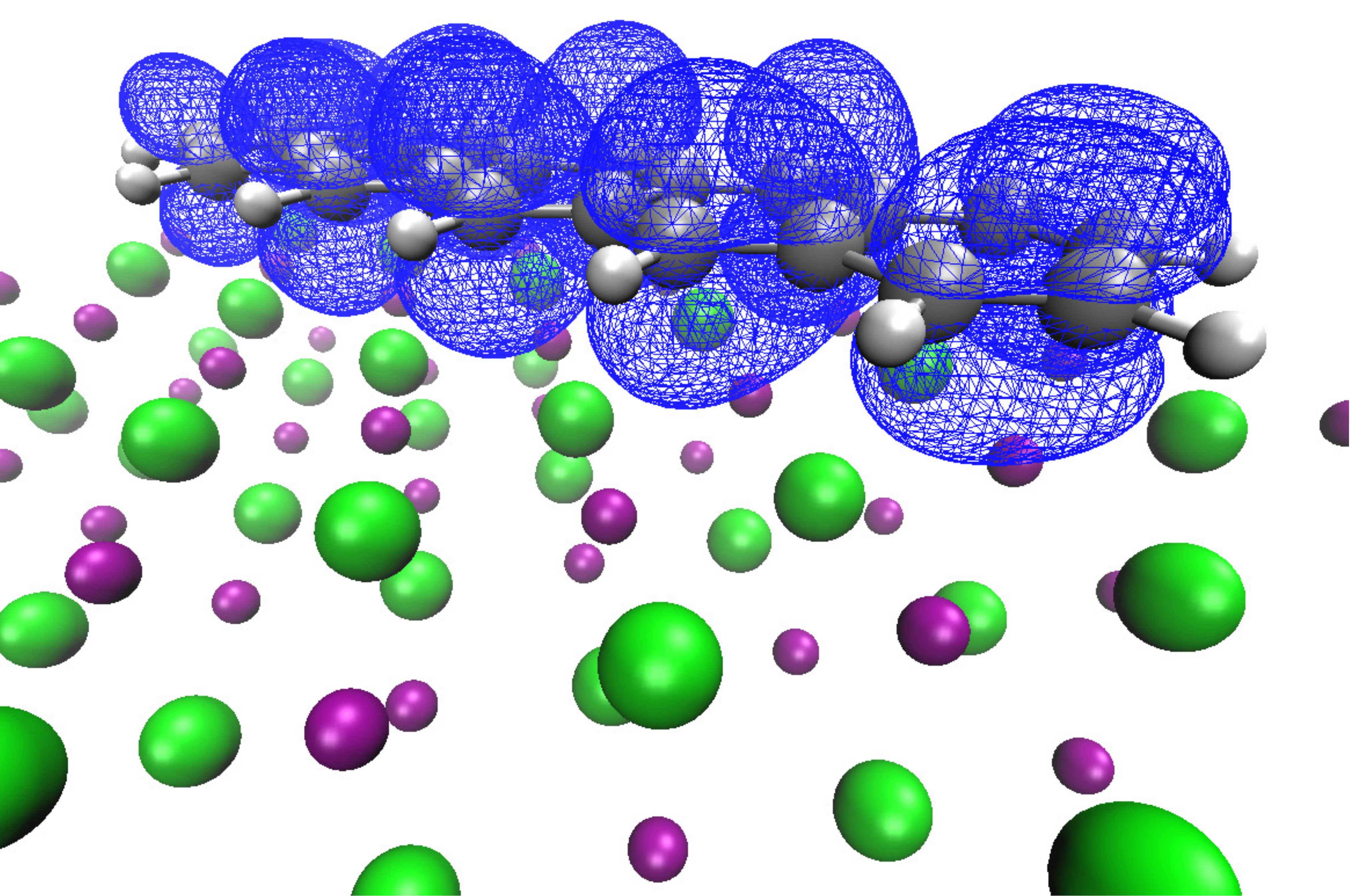}
 &
 \includegraphics[scale=0.13]{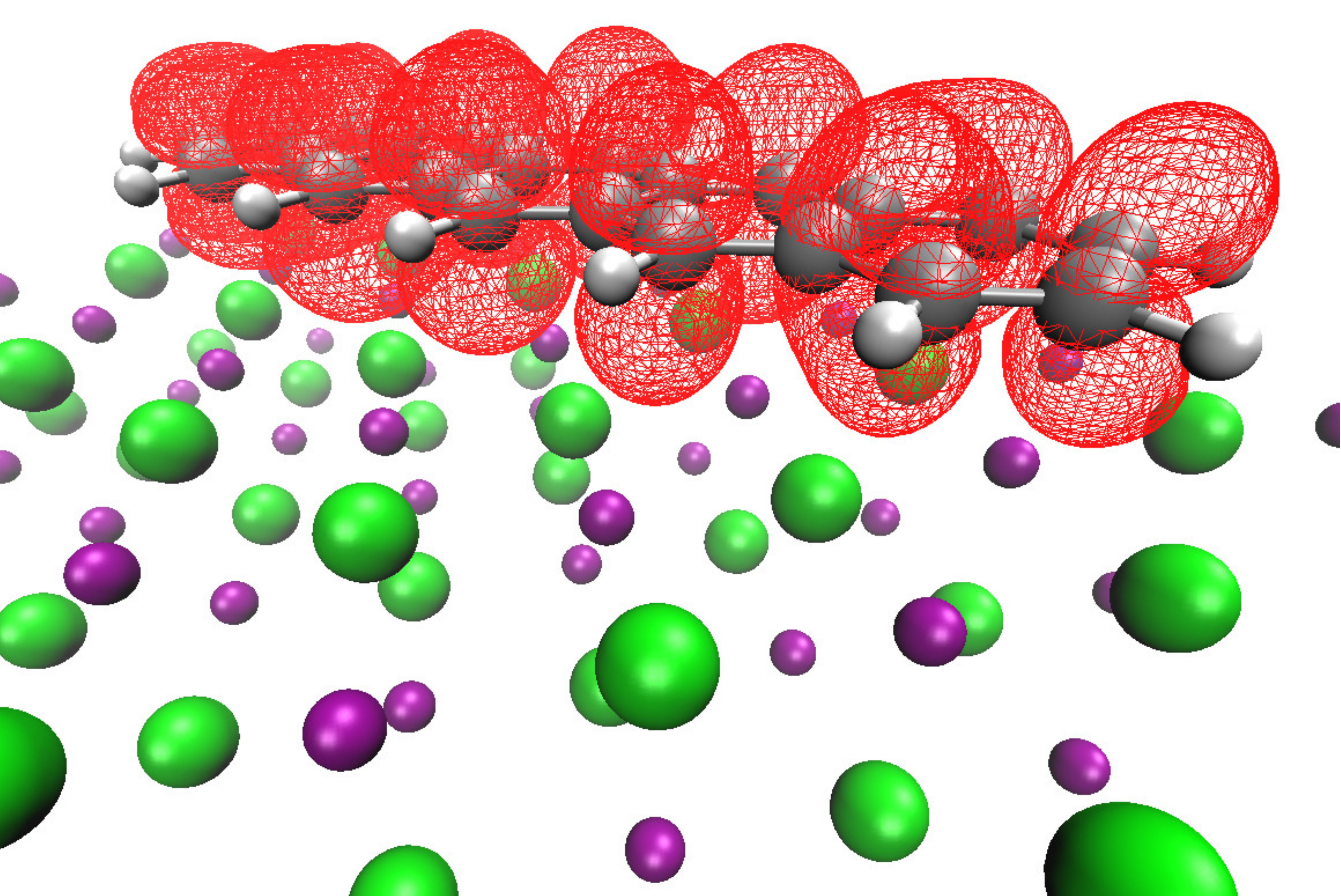}\tabularnewline
 \hline 
\textbf{Cation: SOMO} & \textbf{Anion: SUMO} \tabularnewline
\includegraphics[scale=0.13]{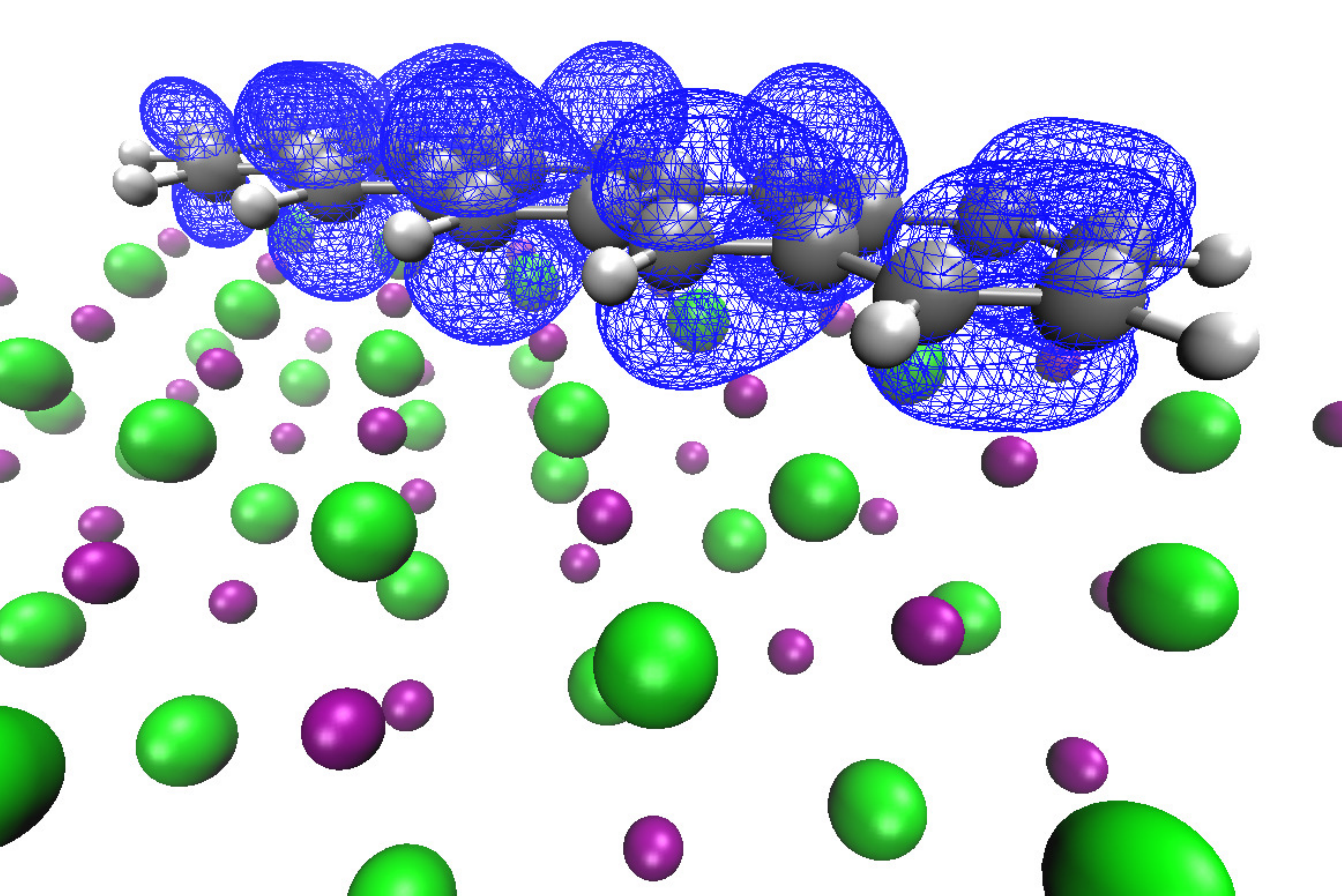}
&
\includegraphics[scale=0.13]{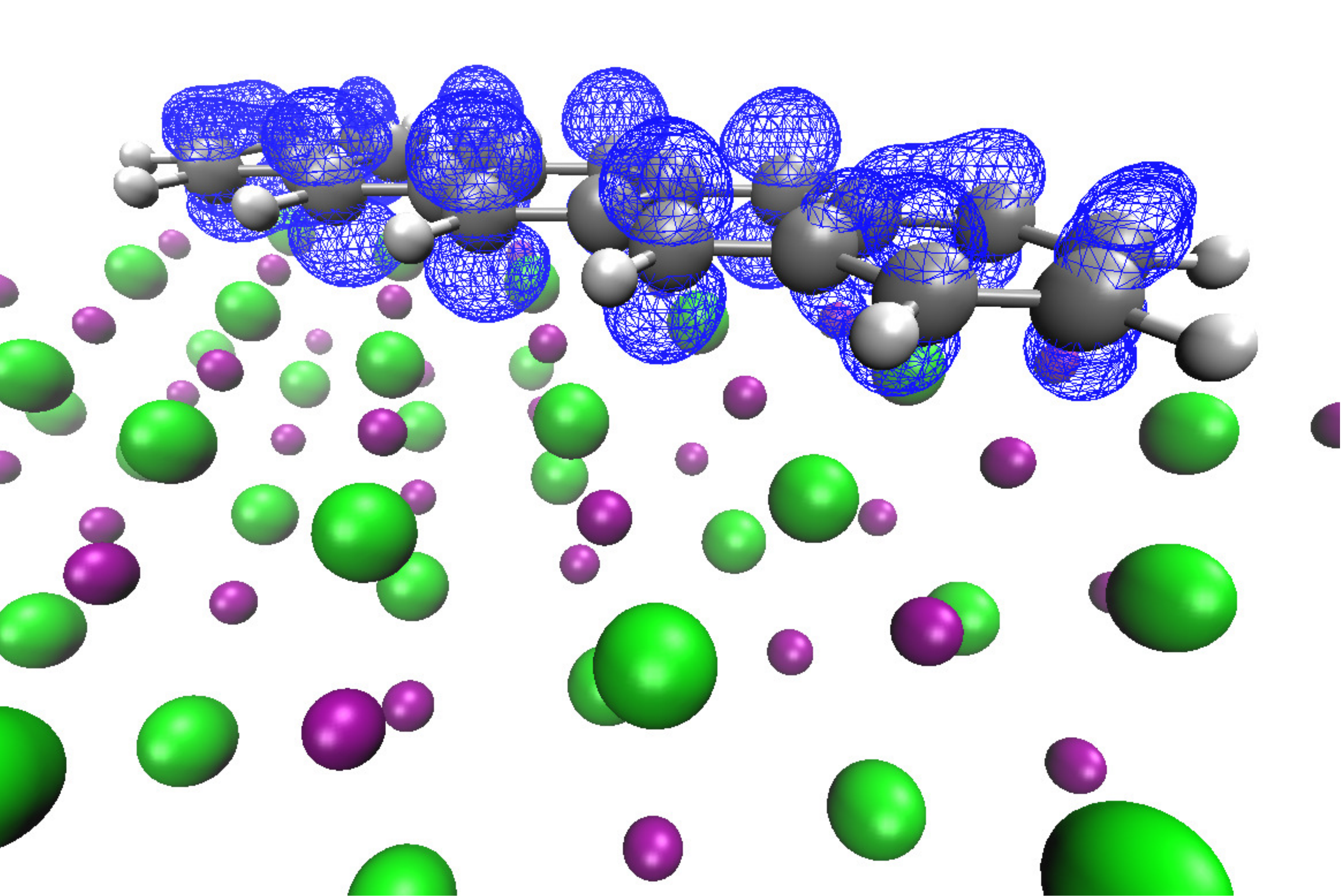}\tabularnewline
\hline 
\end{tabular}
\caption{(online colour) Iso-surfaces of the electron densities of the
  frontier orbitals of the neutral isolated (upper panel) and adsorbed
  (middle panel) pentacene molecule, and of the negatively and
  positively charged adsorbed pentacene molecule (lower
  panel). Results of middle and lower panel were computed with the
  DFT-PC-FF method.  Isosurfaces are taken at $5\times 10^{-3}$
  e/\AA{}$^{3}$. The C and H atoms of pentacene and the Na and Cl ions
  are represented by grey, white, violet and green spheres,
  respectively.}
\label{fig:IR_penta} 
\end{figure}

The electronic structure of absorbed pentacene was analysed using
the PDOS of the C-$p_z$ states of pentacene (the $\pi$ orbitals) and
the $p$ states of the ions in the NaCl bilayer. Results for DFT-FF-PC
and DFT-full are shown in Fig. \ref{fig:pdos_penta}(a) and (b),
respectively. In addition, C-$p_z$ states of isolated pentacene are
shown in Fig. \ref{fig:pdos_penta}(a) as a reference.

A comparison between the PDOS of the DFT-full calculations of adsorbed
pentacene and the DFT calculations of isolated pentacene show that
there is no discernible broadening of of the $\pi$ orbitals due to
their interaction with the states of the film and the substrate. In
addition, the $\pi$ orbitals of adsorbed pentacene, except the orbital
around -6.35 eV, experience only a small rigid downward shift of about
0.30 eV with respect to the vacuum level. The smaller shift of the
orbital at -6.35 eV is due to its interaction with $p$ states of the
ions in the film. the DFT-full results show that the HOMO level  
is 1.09 eV below the Fermi energy ($\varepsilon_{\mathrm{F}}$), whereas
the LUMO is just above $\varepsilon_{\mathrm{F}}$. Finally, the PDOS
of the $p$ states of the ions NaCl in the bilayer shows up as a broad
distribution, which is a result of their interaction with the Cu
substrate states.

A comparison between the upper and middle panels of
Fig.\ref{fig:IR_penta} show that the electron densities of the HOMO
and the LUMO of the adsorbed neutral molecule computed with DFT-PC-FF
are very similar to the corresponding orbital densities of the
isolated molecule and exhibit the same nodal structure. In addition,
the difference between the KS energies gives a HOMO-LUMO gap of 1.17
eV, which is very close to the corresponding KS value of 1.14 eV for
the isolated molecule. From Fig. \ref{fig:pdos_penta}, one finds that
the $p_z$ states of all the C atoms using the DFT-PC-FF method are in
excellent agreement with the DFT-full results obtained by the explicit
inclusion of the Cu(100) substrate. This agreement shows that the
mixing of the $\pi$ orbitals with Cu states are negligible and
provides further support to the application of the DFT-PC-FF method to
this system.

\subsection{Frontier orbitals energy gaps and ionic resonances}
\label{sec:IRres} 
The anion and cation states of the adsorbed pentacene molecule on the
NaCl(2ML)/Cu(100) were calculated using the DFT-PC-FF method by adding
a single electron or hole to the neutral adsorbed pentacene on the
NaCl bilayer at a fixed geometry corresponding to the calculated
equilibrium geometry of the neutral molecule and film. Since the
charging and discharging results in an odd number of electrons, spin
polarisation was included in the simulations. The characters of the
highest occupied orbitals following charging and discharging are
demonstrated by the calculated orbital densities in lower panel of
Fig.~\ref{fig:IR_penta}. These densities correspond to a singly
occupied molecular orbital (SOMO) and a single unoccupied molecular
orbital (SUMO) for the cation and the anion, respectively, which are
very similar in shape to the LUMO and HOMO of the isolated and
absorbed molecule, respectively.

\begin{figure}
\centering \includegraphics[scale=0.35]{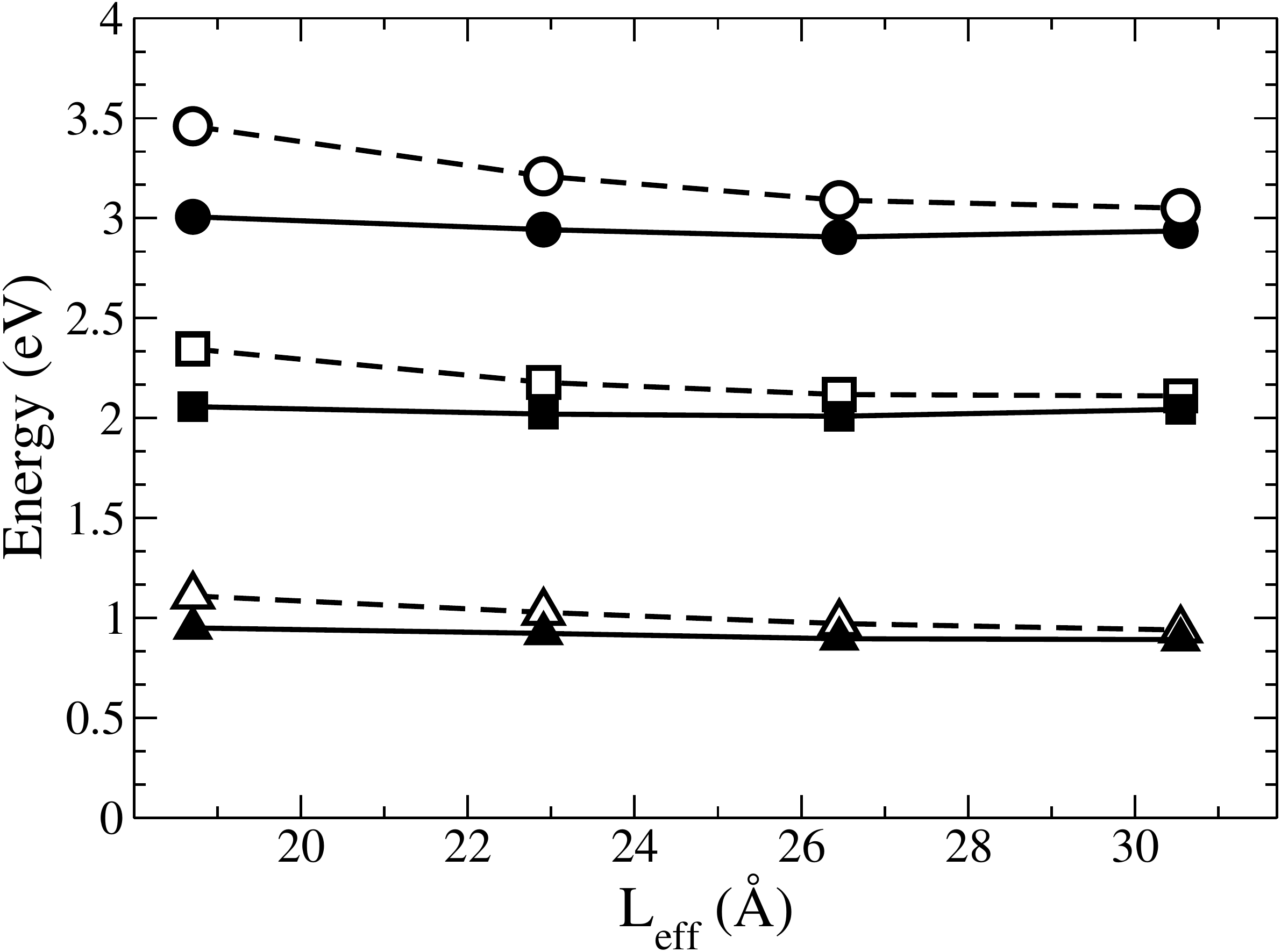}
\caption{Calculated electron (diamonds) and hole (squares) attachment
  energies and HOMO-LUMO energy gaps (circles) for the adsorbed
  pentacene molecule as a function of the effective lateral size of
  the NaCl bilayer $L_\mathrm{eff}$. Due to the electrostatic
  interaction between the infinite replica, computed values clearly
  decrease with $L_\mathrm{eff}$ towards limiting value for zero coverage
  (empty symbols). By substrating the dipole energy correction of
  \ref{app:dipcorr}, convergence to the zero coverage limit is
  much  improved (filled symbols).}
\label{fig:HL_vs_L} 
\end{figure}

In the case of an adsorbed pentacene on a NaCl bilayer,
Fig. \ref{fig:HL_vs_L} shows the calculated vertical electron and hole
attachment energies $\Delta E_\mathrm{e}$ and $\Delta E_\mathrm{h}$ for
different lateral sizes of the supercell or coverage, as obtained from
Eq.(\ref{eq:DE_e_Def}) and (\ref{eq:DE_h_Def}). Here, the effective
lateral size is defined by $L_\mathrm{eff}=\sqrt{L_{x}L_{y}}$ where
$L_{x}$ and $L_{y}$ are the lengths of the supercell along $x$ and $y$
directions, respectively.  Due to long-range electrostatic
interactions between the periodic replica of the adsorbed molecule,
$\Delta E_\mathrm{e}$ and $\Delta E_\mathrm{h}$ converge slowly with
increasing $L_\mathrm{eff}$ to their zero-coverage values of
a single adsorbed molecule. An extrapolation of $\Delta E_\mathrm{e}$
and $\Delta E_\mathrm{h}$ to the zero-coverage limit is done here by
subtracting the dominant long-range electrostatic interaction, the
dipole-dipole interaction, between the periodic replica, as detailed
in \ref{app:dipcorr}. This dipole correction improve considerably the
convergence to the zero-coverage limit, as shown by the dipole-corrected energies $\Delta E_\mathrm{e}$ and $\Delta E_\mathrm{h}$ in
Fig. \ref{fig:HL_vs_L}.  Note that the corresponding dipole-dipole
interactions in the perpendicular $z$ direction are already cancelled
by the dipole correction provided by the dipole layer in the vacuum
region \cite{ISMP_1,ISMP_2}. Using the dipole-corrected energies we
obtain zero-coverage values of 2.06 and 0.95 eV for $\Delta
E_\mathrm{e}$ and $\Delta E_\mathrm{h}$, respectively. The
corresponding HOMO-LUMO gap $E_\mathrm{G}$, as obtained directly from
$\Delta E_\mathrm{e}$ and $\Delta E_\mathrm{h}$ using
Eq. (\ref{eq:egap_Def}), is equal to 3.01 eV. This value is
substantially larger than the value of 1.17 eV obtained from the
calculated KS energies for the HOMO and the LUMO of the neutral
adsorbed molecule. Interestingly, the value of 3.01 eV is a bit less
than the the calculated value of 4.62 eV of $E_\mathrm{G}$ for the
isolated molecule. We attribute this difference to the electronic
polarisation of the positively and negatively charged molecule by the
NaCl film.

The calculated vertical electron and hole attachment energies can be
compared with the experimental energies for the positive and negative
ionic resonances (NIR and PIR) by scanning tunneling spectroscopy of a
single pentacene molecule adsorbed on NaCl films supported by a
Cu(100) substrate \cite{repp2}. A comparison of STM images at the biases
of the NIR and PIR with computed images
showed that these resonances correspond to electron and hole attachment
from the tip to the LUMO and HOMO of the molecule, respectively. The
observed values of 1.3 and -2.8 V for the sample biases of the NIR and
PIR, respectively, give $\Delta E_\mathrm{e}$= 1.3 eV and $\Delta
E_\mathrm{h}=$2.8 eV and a value of 4.1 eV for $E_\mathrm{G}$.
Thus, the experimental value of $E_\mathrm{G}$ is
reduced by approximately 1.17 eV upon adsorption on the bilayer.
Surprisingly, the deviation of about 1.09 eV between the
experimental (4.1 eV) and computed value (3.01 eV) for $E_\mathrm{G}$
is somewhat a bit larger than the corresponding deviation of 0.65 eV
for the isolated molecule in vacuum (5.27 eV and 4.62 eV for
experimental and computed $E_\mathrm{G}^{0}$, respectively, as
reported in section \ref{subsec:isomol}).

\begin{figure}
\centering \includegraphics[scale=0.5]{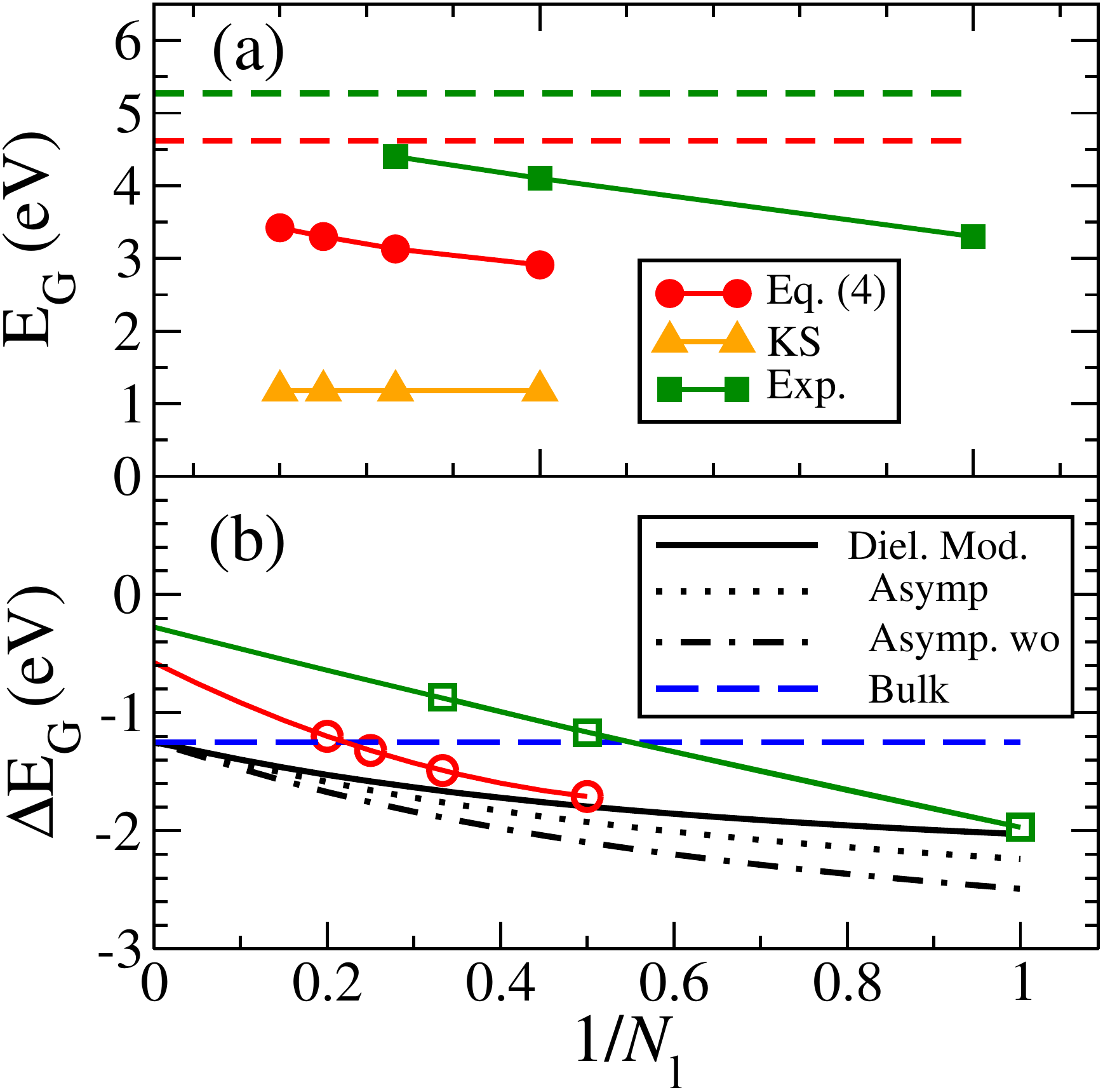}
\caption{(online colour) (a) Calculated and experimental HOMO-LUMO
  energy gaps, $E_\mathrm{G}$, for the adsorbed pentacene molecule and
  (b) its shift $\Delta E_\mathrm{G}$ with respect to its gas-phase
  value as a function of $1/N_\mathrm{l}$, where $N_\mathrm{l}$ is the
  number of NaCl monolayers. The calculated $E_\mathrm{G}$ (solid
  circles) were obtained from DFT-PC-FF calculations of the negatively
  and positively charged molecules using Eq. (\ref{eq:egap_Def}),
  including lateral dipole corrections. (a) $E_\mathrm{G}$ obtained
  from the calculated Kohn-Sham energies are significantly
  underestimated (solid triangles). The experimental $E_\mathrm{G}$
  (solid squares) were obtained from measurements by scanning
  tunneling spectroscopy \cite{repp2}. Note that the uncertainities in
  the experimental values are several tenths of an eV~\cite{uncertain}. The
  calculated and the experimental $E_\mathrm{G}$ of the isolated
  molecule are also indicated by red and green dashed lines,
  respectively.  (b) The extrapolation of the calculated and the
  experimental values of $\Delta E_\mathrm{G}$ to
  $N_\mathrm{l}\rightarrow\infty$ was obtained by a fit to a quadratic
  polynomial in $1/N_\mathrm{l}$.  We have also included the results
  for $\Delta E_\mathrm{G}$ in the dielectric model,
  Eq.(\ref{eq:EgResDielMod}), (solid line) and also its asympotic
  result for $\Delta E_\mathrm{G}$, Eq.(\ref{eq:DEGAsymp}), with
  (dotted line) and without (dot-dashed line) the dielectric screening
  of image interaction with the perfect conductor. The bulk limit of
  $\Delta E_\mathrm{G}$ in the dielectric model is indicated by the
  blue dashed line.}
\label{fig:HL_vs_layers} 
\end{figure}

Here, the effect of the NaCl film on $E_\mathrm{G}$ was investigated
by calculating $E_\mathrm{G}$ as a function of the number of
monolayers $N_l$ of the film. In these calculations, the values of
$E_\mathrm{G}$ from 2 to 5 ML were obtained from the dipole-corrected
$\Delta E_\mathrm{e}$ and $\Delta E_\mathrm{h}$ for a supercell with
$4\times3$ surface unit cell (see Section \ref{sec:comp_methods})
corresponding to $L_{eff}\approx$26.5 \AA{}.  Computed values of
$E_{\mathrm{G}}$ as a function of $1/N_l$ are shown in
Fig.\ref{fig:HL_vs_layers} (a). In addition, we have included the
calculated gap from the KS energies as well as the experimental values
taken from Ref.\cite{repp2}. For a comparison, we also show in dashed
lines computed and experimental values of the isolated molecule. The
observed increase of $E_\mathrm{G}$ with increasing $N_l$ is
reproduced by calculated values. In particular, the observed reduction
of about 0.3 eV for the experimental values of $E_\mathrm{G}$ between
the bilayer ($1/N_\mathrm{l}=1/2$) and the trilayer
($1/N_\mathrm{l}=1/3$) is rather well reproduced by the calculated
reduction of 0.22 eV.

The increase of $E_\mathrm{G}$ with $N_l$ can be understood in a
simple dielectric model from the screening by the metal substrate and
the film, as detailed in \ref{app:DielMod}. The metal is modelled by a
perfect conductor, whereas the film is modelled by an homogeneous
dielectric with an effective thickness $a(N_\mathrm{l})$ and
electronic dielectric constant $\epsilon_{\mathrm{\infty}}$ (see
Fig. \ref{fig:dielmodel}).  The values of these parameters and the
distance $z_{\mathrm{d}}-z_{\mathrm{s}}$ between the dielectric vacuum
interface and the average NaCl surface layer position $z_{s}$ were
determined by applying an homogeneous external electric field to the
adsorbed film, as detailed in \ref{app:DielPar}.  The electronic
dielectric constant $\epsilon_{\mathrm{\infty}}$ and
$z_{\mathrm{d}}-z_{\mathrm{s}}$ were both found to be essentially
independent of $N_\mathrm{l}$ with $\epsilon_{\mathrm{\infty}}\approx$
2.64 and $z_{\mathrm{d}}-z_{\mathrm{s}}\approx$ 1.66 {\AA}, while the
effective thickness of the film was well-approximated by
$a(N_\mathrm{l})\approx a_{0}+N_\mathrm{l}\Delta a$ for
$N_\mathrm{l}=2-5$ where $a_{0}\approx 0.84$ {\AA} and $\Delta
a\approx$ 2.81 {\AA}. The value for $\Delta a$ is essentially equal to
the interlayer distance. In addition, the lateral extension of the
surface charge distribution $\sigma\mathrm{_{ext}}({\bf R})$ of the
positively and negatively charged molecule has simply been modelled as
an homogeneously charged rectangular sheet with a net charge of $\pm e$.
The values for the side lengths $D_{x}=$ 14.1 {\AA} and $D_{y}=$ 5.0
{\AA} of the sheet were simply determined from the length and width of the
pentacene molecule.

In this model, the layer dependence of the energy difference $\Delta
E_{G}(N_\mathrm{l})$ between the energy gap for the adsorbed and
isolated molecule is given by (See \ref{app:DielMod})
\begin{eqnarray}
\Delta E_{G}(N_\mathrm{l}) & = & E_{G}(N_\mathrm{l})- E_{G0} \\
& = & -\int\int\frac{d^{2}K}{2\pi
  K}\frac{(\epsilon_{\mathrm{\infty}}-1)\exp(2Ka(N_\mathrm{l}))+(\epsilon_{\mathrm{\infty}}+1)}{(\epsilon_{\mathrm{\infty}}+1)\exp(2Ka(N_\mathrm{l}))
  +(\epsilon_{\mathrm{\infty}}-1)}|\sigma\mathrm{_{ext}}({\bf
  K})|^{2}\exp(-2KD) \nonumber \label{eq:EgResDielMod}
\end{eqnarray}
where $D$ is the distance of the molecule from the dielectric-vacuum
interface and the lateral Fourier transform of the charged sheet
$\sigma\mathrm{_{ext}}({\bf K})$ is given by
Eq.(\ref{eq:rhoextFourier}). Here, $D$ is determined from the
calculated equilibrium distance $d=3.06$ {\AA} of the molecule from
the outermost NaCl layer, as $D=d-z_{d}+z_{s}$=1.36 {\AA}. The
resulting $\Delta E_{G}(N_\mathrm{l})$ is shown in
Fig. \ref{fig:HL_vs_layers}(b). A better understanding of the
$N_\mathrm{l}$ dependence $\Delta E_{G}(N_\mathrm{l})$ in this model
is obtained from an asymptotic expansion for large $N_l$, which is
given by (see, \ref{app:DielMod}),
\begin{equation}
\Delta E_{G}(N_\mathrm{l}) \asymp
-\frac{\epsilon_{\infty}-1}{2(\epsilon_{\infty}+1)}\frac{\tilde{Q}_\mathrm{F}^{2}}{D} 
-\frac{4\epsilon_{\infty}}{2(\epsilon_{\infty}+1)^2}\frac{\tilde{Q}_\mathrm{M}^{2}}{D+N_\mathrm{l}a}, \ N_\mathrm{l} \rightarrow \infty .
\label{eq:DEGAsymp}
\end{equation}
Here, the effective charges $\tilde{Q}_{F,M}$ take into account the
lateral extension of the charge distribution of the charged molecule
and are defined in Eq.~(\ref{eq:QeffDef}). In the case of a point
charge with charge $\pm e$,
$\tilde{Q}^2_\mathrm{F}=\tilde{Q}^2_\mathrm{M}=e^2$ but
$\tilde{Q}_\mathrm{F,M}$ decreases for an increasing lateral extension
of the charge distribution. In the case of our simple model for the
charge distribution of the HOMO and LUMO of pentacene the reduction is
quite substantial $\tilde{Q}^2_\mathrm{F}=0.53e^2$ and but much less
so for $\tilde{Q}^2_\mathrm{M}=0.92-0.96e^2$ due to the much larger
distance of the molecule from the metal surface than its distance from
the dielectric film. The leading order term on the right hand side of
Eq.~(\ref{eq:DEGAsymp}) gives the energy gap $\Delta
E_{G}(N_\mathrm{l}\rightarrow\infty)$ of the molecule adsorbed on a
bulk dielectric, whereas the second term gives the contribution to
$E_{G}(N_\mathrm{l})$ from the image interaction with the metal
surface screened by the dielectric film. The asymptotic result in
Eq.(\ref{eq:DEGAsymp}) for $E_{G}(N_\mathrm{l})$ is also shown in in
Fig. \ref{fig:HL_vs_layers} (b) and is close to the full result from
Eq.~(\ref{eq:EgResDielMod}). Here, we have also illustrated that the
dielectric screening of the image interaction with the metal gives
rise to relatively small reduction of
$\frac{(\epsilon_{\infty}+1)^2}{4\epsilon_{\infty}}\approx 1.25$ in
this case by showing the corresponding result for the unscreened image
interaction with the surface.

As shown in Fig. \ref{fig:HL_vs_layers} (b), from a comparison of the
results of the dielectric model for $\Delta
E_\mathrm{G}(N_\mathrm{l})$ with the calculated results using
DFT-PC-FF, the bulk limit of $\Delta
E_\mathrm{G}(N_\mathrm{l}\rightarrow \infty)$ is severly overestimated
by the dielectric model. The dielectric model gives a value of -1.25
eV in this limit for $\Delta E_\mathrm{G}(N_\mathrm{l})$ whereas an
extrapolation of the calculated values to $N_\mathrm{l}\rightarrow
\infty$ gives a value of only about -0.58 eV for the downward
shift. The corresponding extrapolated value from the experiments gives
even a smaller downward shift of -0.28 eV but might not be a significant
difference due to the uncertainties in experimental
values~\cite{uncertain}. The significant overestimate of the downward
shift in the bulk limit of $\Delta E_\mathrm{G}$ in the dielectric
model demonstrates the challenges of modelling the response of an
ionic insulating film to a charged adsorbate at a close distance where
the electron density distributions of the dielectric film and the
charged molecule are not well-separated. The variation of $\Delta
E_\mathrm{G}$ obtained from DFT-PC-FF is better described with the
dielectric model than the bulk limit $\Delta
E_\mathrm{G}(N_\mathrm{l}\rightarrow \infty)$ but tends to
underestimate this variation. A somewhat surprising behaviour of the
measured $\Delta E_\mathrm{G}(N_\mathrm{l})$ is its near-linear
behaviour even down to $N_\mathrm{l} = 1$, which is not really
captured either by the DFT-PC-FF calculation or the dielectric
model. However, this difference in behaviour might not be significant
due to the uncertainities in the experimental values~\cite{uncertain}.

\section{Summary and conclusions}
\label{sec:conclusions}
In this work, we have addressed the problem of calculating electron
and hole attachment energies of an adsorbed molecule, whose electronic
states are essentially decoupled from the conduction electron states
of the metal substrate by an insulating film. To this purpose, we have
used our recently developed DFT-PC-FF method, where both the film and
the adsorbed molecule are treated fully within DFT, whereas the metal
substrate is treated implicitly by a perfect conductor (PC) model. The
remaining non-Hartree interactions between the metal substrate and the
film are modelled by a simple force field (FF) whose parameters are
obtained from DFT calculations. 

As an example case, we have considered a pentacene molecule adsorbed
on NaCl films supported by a Cu(100) surface and compared our results
with scanning tunneling spectroscopy and atomic force microscopy
experiments. Support for our method comes from the very good agreement
of the DFT-PC-FF results for the relaxed geometry and adsorption
energy of the adsorbed molecule on an NaCl bilayer supported by a
Cu(100) surface with the results from our DFT calculations which
include the metal substrate explicitly. The adsorbed molecule is found
to be neutral and keeps the planar geometry of gas phase. The molecule
is physisorbed on the film as evidenced by the calculated PDOS which
shows that the frontier orbitals experience only a small rigid shift
in energy.

The calculated HOMO-LUMO energy gap for a fixed geometry increases
with the number of NaCl layers in agreement with the experiments. The
calculated energy gap underestimate somewhat the experimental gap, in
part due to the underestimate of the observed gap of the isolated
molecule by the used exchange-correlation energy functional. The
calculated energy gap is much improved over the energy gap as obtained
from  the calculated KS energies for the HOMO and the LUMO.

The layer dependence of the calculated energy gap was analyzed in a
simple dielectric model of the adsorbed film with parameters taken
from DFT calculations of the response of the adsorbed film to an external
electric field. This model rationalizes semi-quantitatively the
observed behaviour of the energy gap with the number of layers and
elucidates the contributions from the film and the metal substrate to
the shift of the energy gap with the number of layers. In particular,
this model reveals that the decrease of the energy gap with decreasing
number of layers is primarily due to the electrostatic interaction with the
metal. Nevertheless, values calculated with this model depart
significantly from the experimental and the calculated DFT-PC-FF values. This
highlights the challenges of building a proper electrostatic model of
charged adsorbates on a insulating film.

\ack The authors acknowledge Leverhulme Trust for funding this project
trough the grant (F/00 025/AQ) and allocation of computer resources
at HECToR through the membership in the materials chemistry consortium
funded by EPSRC (EP/L000202F) and at Lindgren PDC through SNIC. We
also thank Prof. Jasha Repp and Dr. Gerhard Meyer for useful input. Mats
Persson is grateful for the support from the EU project ARTIST.

\appendix

\section{Dipole-dipole energy correction from the interaction between periodic
images}
\label{app:dipcorr} 

In periodic DFT calculations the adsorbate coverage is determined by
the lateral size of the supercell. In order to study single adsorbates
corresponding to the zero coverage limit one needs to perform
calculations for increasing lateral sizes of the supercell in
order to extrapolate the results to the limit of infinite lateral size
or zero coverage. This task can be computationally very demanding due
to the slow convergence caused by long-range electrostatic
interactions of the adsorbate with its periodic images. The
dipole-dipole interaction term is the leading order term of these
interactions. Here we provide an explicit expression
for lateral dipole-dipole energy correction in the PC model for
charged adsorbates which significantly improves the convergence to the
zero-coverage limit. Note that the perpendicular dipole-dipole
interactions are already corrected for by the introduction of a dipole
layer in the vacuum region.

Following the notation introduced in Ref. \cite{ISMP_1} for the PC
model $\rho_\mathrm{s}({\bf r})$ is the charge density of the charged
system that includes both the insulating film and the charged
adsorbate inside a supercell $V$ and $\rho_\mathrm{ind}({\bf r})$ is
the induced charge at the PC plane which is located at the image
position $z_{{\rm im}}$ of the bare metal substrate . The resulting
electrostatic potential $\phi({\bf r})$ has not only a contribution
from $\rho({\bf r})=\rho_\mathrm{s}({\bf r})+\rho_{{\rm ind}}({\bf
  r})$ but also a contribution from a dipole layer
$\rho_\mathrm{dip}(z)$ at a plane in the vacuum region which
compensates for the perpendicular dipole-dipole
interactions. The undetermined constant of $\phi({\bf r})$ is fixed by the condition that $\phi({\bf
  r})=0$ inside the metal. The corresponding charge density and
electrostatic potential in the absence of the charged adsorbate is
denoted by $\rho_\mathrm{s0}({\bf r})$ and $\phi_{0}({\bf r})$
respectively. The change in electrostatic interaction energy upon
adsorption is then given by
\begin{eqnarray}
\Delta E_{{\rm el}}=\frac{1}{2}\intop_{V}\rho_\mathrm{s}({\bf r})\phi({\bf r})d^{3}r-\frac{1}{2}\intop_{V}\rho_{s0}({\bf r})\phi_{0}({\bf r})d^{3}r \label{eq:DEelDef} \\
=\frac{1}{2}\intop_{V}\Delta\rho_\mathrm{s}({\bf r})\phi_{0}({\bf r})d^{3}r+
\frac{1}{2}\intop_{V}\rho_{s0}({\bf r})\Delta\phi({\bf r})d^{3}r
+\frac{1}{2}\intop_{V}\Delta\rho_\mathrm{s}({\bf r})\Delta\phi({\bf r})d^{3}r
\label{eq:DEel}
\end{eqnarray}
where $V$ is the volume of the supercell $\Delta\rho_\mathrm{s}({\bf r})=\rho_\mathrm{s}({\bf
  r})-\rho_\mathrm{s0}({\bf r})$ and $\Delta\phi({\bf r})=\phi({\bf
  r})-\phi_{0}({\bf r})$. The first term 
in Eq.~(\ref{eq:DEel}) is the electrostatic potential energy of the
localized adsorbate-induced charge density $\Delta\rho_\mathrm{s}({\bf
  r})$ in the potential $\phi_{0}({\bf r})$ and will be rapidly
convergent with increasing lateral size of the supercell. The two remaining terms in Eq.~(\ref{eq:DEel}) can be handled
by introducing the electrostatic field $\Delta\phi^{(0)}({\bf r})$
from the charge distribution induced by the adsorbate in the zero-coverage limit
\begin{equation}
\Delta\phi^{(0)}({\bf r}) = 
\int \frac{\Delta\rho({\bf r}^\prime)}{|{\bf r}-{\bf r}^\prime|}d^{3}r^\prime \ .
\end{equation}
The potential $\Delta\phi({\bf r})$ can now be expressed in terms of $\Delta\phi^{(0)}({\bf r})$ as a sum over the lateral lattice vectors ${\bf R}_\|$ as
\begin{equation}
\Delta\phi({\bf r}) = \sum_{{\bf R}_\|}\Delta\phi^{(0)}({\bf r}-{\bf R}_\|) \ .
\label{eq:PhiDecomp}
\end{equation}
Note that it is sufficient to restrict the summation over ${\bf R}_\|$ due to the inclusion  of the dipole layer $\rho_\mathrm{dip}(z)$ in the supercell.
Using this decomposition and the periodicity of $\rho_\mathrm{s0}({\bf r})$ the second term in Eq.~(\ref{eq:DEel}) is given by
\begin{equation}
\frac{1}{2}\intop_{V}\rho_{s0}({\bf r})\Delta\phi({\bf r})d^{3}r = 
\frac{1}{2}\int\rho_{s0}({\bf r})\Delta\phi^{(0)}({\bf r})d^{3}r 
\end{equation}
which is nothing else than its zero coverage limit. Thus only the
third term on the RHS in Eq.~(\ref{eq:DEel}) contains the long-range
electrostatic interactions and is given in term of the decomposition
in Eq.~(\ref{eq:PhiDecomp}) as
\begin{equation}
\frac{1}{2}\intop_{V}\Delta\rho_\mathrm{s}({\bf r})\Delta\phi({\bf r})d^{3}r =
\frac{1}{2}\sum_{{\bf R}_\|} \intop_{V}\Delta\rho_\mathrm{s}({\bf r})\Delta\phi^{(0)}({\bf r}-{\bf R}_\|)d^3r
\label{eq:DRhosDPhiRes}
\end{equation}
The ${\bf R}_\|={\bf 0}$ term converges rapidly to the zero-coverage
limit and the remaining terms which are the electrostatic potentials
from the periodic images can be approximated by a multipole
expansion. The leading order term in this expansion of the
electrostatic potential of the neutral charge distribution
$\Delta\rho({\bf r})$ will be a dipole potential given by
\begin{equation}
\Delta\phi^{(0)}({\bf r}-{\bf R}_\|) \simeq \frac{2\Delta\mu_z(z-z_{{\rm im}})}{R_\|^3} \ .
\label{eq:DipPotRes}
\end{equation}
Note that the screening by the perfect conductor gives rise to a dipolar electrostatic field with an effective dipole moment which is twice as
large as the dipole moment $\Delta\mu_z$ of the adsorbate-induced charge distribution given by
\begin{equation}
\Delta\mu_{z}=\intop_{V}z\Delta\rho({\bf r})d^{3}r=\intop_{V}(z-z_{{\rm im}})\Delta\rho_{s}({\bf r})d^{3}r \ .
\label{eq:DmuzDef}
\end{equation}
From Eqs.(\ref{eq:DRhosDPhiRes}),(\ref{eq:DipPotRes}) and (\ref{eq:DmuzDef}) one obtains that the electrostatic interaction energy 
$\Delta E^\prime_\mathrm{el}$ from the periodic images is given by
\begin{equation}
\Delta E_{\rm el}^\prime \simeq 
\Delta\mu_{z}^{2}\sum_{{\bf R_\|}\neq{\bf 0}}\frac{1}{R_{\|}^{3}}\ .
\label{eq:DEelpRes}
\end{equation}
This interaction energy is repulsive and should be subtracted from the calculated total energies in order to correct for the dipole-dipole interactions between the periodic images. The lattice sum in Eq.~(\ref{eq:DEelpRes}) can be readily evaluated numerically. For a supercell with a square lateral shape of side length $L$ this interaction energy decays as $L^{-3}$

Finally note that this result as applied to an external charge
outside the PC differs from the result that would be obtained from
Eqs. (59) (62) and (63) in Ref. \cite{ISMP_1} in one important
aspect. The results in Eqs. (59) and (62) were obtained from the
electrostatic potential $\phi_{{\rm ind}}({\bf r})$ of $\rho_{{\rm
    ind}}({\bf r})$ rather than from $\phi({\bf r})$ as done
here. Thus the result in Ref. \cite{ISMP_1} contains a contribution
that decays as $L^{-1}$ as first pointed out by G. Makov and
M. C. Payne \cite{makov}.

\section{A simple dielectric model of the adsorbed film}
\label{app:DielMod}

\begin{figure}
\centering
\includegraphics[scale=1.0]{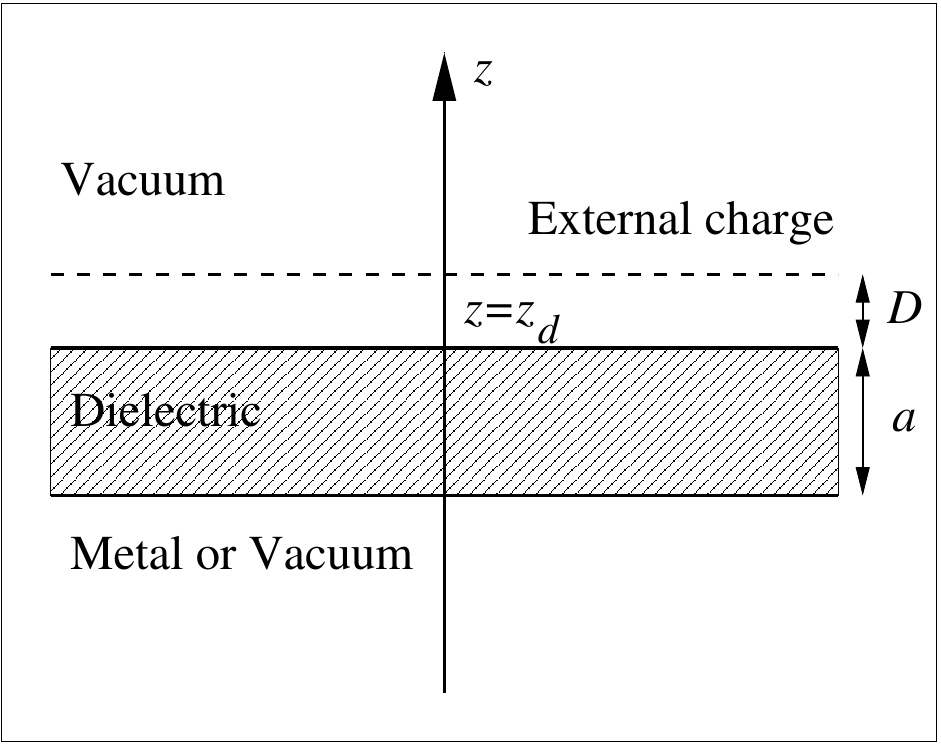}
\caption{Schematics of the dielectric model of an
adsorbed molecule on an ionic insulating film, which is either freestanding
or supported by a perfect conductor model of the metal substrate. The dashed line indicates the position of the externally charged sheet.}\label{fig:dielmodel}
\end{figure}

Here,  we will derive the interaction energy for an external surface
charge distribution outside a dielectric film supported by a perfect
conductor (PC) model of the metal substrate.  In this model, the shift
of the HOMO-LUMO energy gap from its value for an isolated molecule is
then given by the sum of the corresponding electrostatic interaction
energies of the charge distributions for both the positively and
negatively charged molecule with the supported film. As illustrated schematically in
Fig. \ref{fig:dielmodel}, the metal is modelled by a PC and the
adsorbed film by a homogeneous dielectric film with thickness $a$ and
dielectric constant $\epsilon$. Here we will use the notation ${\bf
  r}=({\bf R},z)$.

The interaction energy of an external surface charge distribution
$\sigma_\mathrm{ext}({\bf R})$ at a distance $D$ from the dielectric film
is given by,
\begin{equation}
E_\mathrm{int}=-\frac{1}{2}\int\int\frac{d^{2}K}{2\pi K}r(K)|\sigma_\mathrm{ext}({\bf K})|^{2}\exp(-2KD)\label{eq:EintRes}
\end{equation}
where $\sigma_\mathrm{ext}({\bf K})$ is the lateral Fourier transform
of $\sigma({\bf R})$ and $|{\bf K}|=K$. $r(K)$ is the reflection
coefficient from the dielectric-vacuum interface at $z=z_{d}$ of the
evanescent plane wave component $\phi_\mathrm{ext}({\bf K})\exp(Kz)$
of the external electrostatic potential from $\sigma_\mathrm{ext}({\bf R})$. In
the region $z< z_d+D$, the plane wave component $\phi(z;{\bf K})$ of
the electrostatic potential is then given by
\begin{equation}
\label{eq:phiRes}
\phi(z;{\bf K})=\phi_{\mathrm{ext}}({\bf K})
\begin{cases}{
\exp(K(z-z_{d}))-r(K)\exp(-K(z-z_{d}))  \, \,  z_d+D>z>z_{d} \\
t(K)\exp(K(z-z_{d}))-s(k)\exp(-K(z-z_{d}))  \, \,  z_{d}>z>z_{d}-a\\
0  \, \,  z_{d}-a>z}
\end{cases}
\end{equation}
The reflection coefficient $r(K)$ and the coefficients $s(K)$ and $t(K)$ in
Eq. (\ref{eq:phiRes}) can now be determined from the boundary conditions
that the parallel component of the electric field and the perpendicular
component of the external electric field should be both continuous
across the two interfaces. These boundary conditions gives, 
\begin{equation}
r(K)=\frac{(\epsilon-1)\exp(2Ka)+(\epsilon+1)}{(\epsilon+1)\exp(2Ka)+(\epsilon-1)}.
\label{eq:rRes}
\end{equation}
In the asympotic limit of a thick dielectric film corresponding to
$a\rightarrow\infty$, $r(K)$ in Eqns.(\ref{eq:rRes})
reduces to,
\begin{equation}
r(K) \asymp \frac{\epsilon-1}{\epsilon+1} + \frac{4\epsilon}{(\epsilon+1)^2}\exp(-2Ka)
\label{eq:rResAsymp}
\end{equation}

Here we will go beyond the simple point charge model for the charge
distribution of an adsorbed molecule with charge $Q$ and take into
account of the lateral extension of this charge distribution by using
the following simple rectangular surface charge distribution 
\begin{equation}
\sigma\mathrm{_{ext}}({\bf R})=\begin{cases}{
\frac{Q}{D_x D_y}  |X|<\frac{D_x}{2}\ \mathrm{and}\ |Y|<\frac{D_y}{2}\\
0 \, \, \, \, \, \, \, \, \mathrm{otherwise}}
\end{cases}\label{eq:rhoextModel}
\end{equation}
whose Fourier transform is given by 
\begin{equation}
\sigma_\mathrm{ext}({\bf K})=\frac{4Q\sin\left(\frac{D_xK_x}{2}\right)\sin\left(\frac{D_yK_y}{2}\right)}{D_xK_xD_yK_y}\label{eq:rhoextFourier}
\end{equation}
The corresponding $E_\mathrm{int}$ are then readily calculated
from Eqs.(\ref{eq:EintRes}) thanks to the exponential decay of the
integrand with $K$ using a two-dimensional numerical quadrature .

The asymptotic interaction energy as obtained from $r(K)$ in
Eqs.(\ref{eq:rResAsymp}) and (\ref{eq:rhoextFourier}) can now be expressed as,
\begin{equation}
E_\mathrm{int} \asymp -\frac{\epsilon-1}{4(\epsilon+1)}\frac{Q^{2}}{D}
- \frac{4\epsilon}{(\epsilon+1)^2}\frac{Q^{2}}{4(D+a)},
\label{eq:EintResAsymp}
\end{equation}
where the effective charges $\tilde{Q}_{F,M}$ are given by,
\begin{equation}
\tilde{Q}^{2}_\mathrm{F,M}=\int\int\frac{d^{2}K}{2\pi
  K}|\sigma\mathrm{_{ext}}({\bf
  K})|^{2}2h_\mathrm{F,M}\exp(-2Kh_\mathrm{F,M}).
\label{eq:QeffDef}
\end{equation}
Here, $h_\mathrm{F}=D$ and $h_\mathrm{M}=D+a$ are the
distances between the external charge and the surface of the
dielectric film and the perfect conductor, respectively.  Note that
the first term in (\ref{eq:EintResAsymp}) is the interaction energy of the
external charge distribution with a semi-infinite dielectric and the
second term reduces to the corresponding interaction energy with a
perfect conductor in the absence of the dielectric film $\epsilon =
1$. Thus the prefactor $\frac{4\epsilon}{(\epsilon+1)^2}$ in the
second term is due to the dielectric screening by the film.

\section{Determination of dielectric parameters for the adsorbed film}
\label{app:DielPar}

\begin{figure}
\centering 
\includegraphics[scale=0.3]{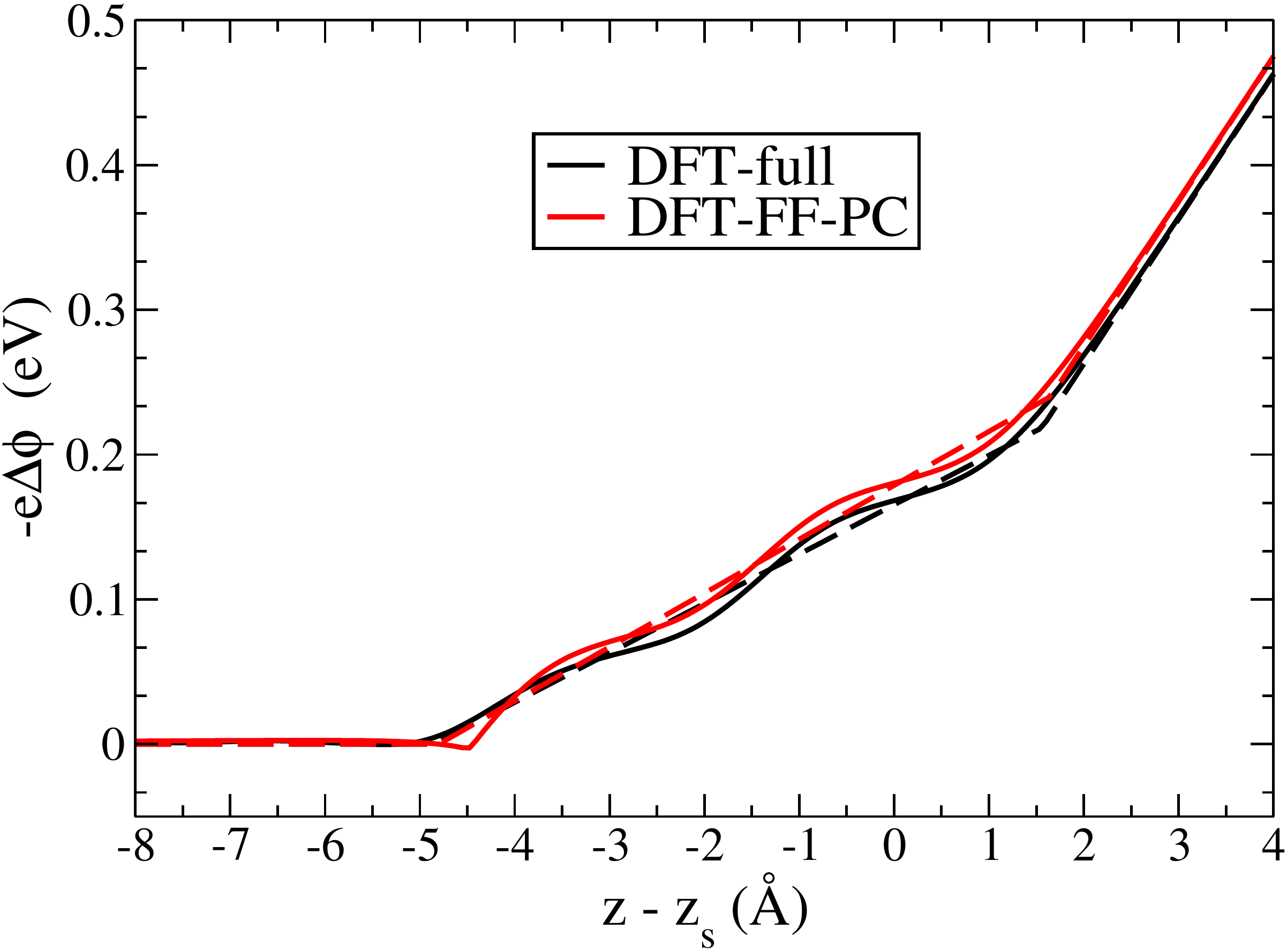}
\caption{(online color) Calculated electrostatic potential energy induced by an external perpendicular homogeneous electric field
  across a NaCl bilayer supported by a Cu(100) surface as obtained by
  DFT (black solid line) and DFT-FF-PC (red solid line) for fixed
  nuclear positions. The dashed lines are corresponding least square
  fits to the dielectric model electrostatic potential in
  Eq. \ref{eq:PotDielMod}. $z_s$ is the position of the outermost NaCl plane. The strength of the external electric field
  is 0.05 V/{\AA}.}
\label{fig:indpotcomp} 
\end{figure}

Here we show how the effective thickness $a$ of the ionic insulating
film and its effective electronic dielectric constant
$\epsilon_\infty$ were determined from the calculated response of the
adsorbed film to an external homogeneous electric field ${\bf
  E}_\mathrm{ext}=E_\mathrm{ext}\hat{z}$ for fixed nuclear
positions. This electric field was included in our DFT-full and
DFT-PC-FF calculations using the method described in
Ref.~\cite{ISMP_2}. Results for the induced electrostatic potentials
$\Delta\phi(z)$ are shown in Fig. \ref{fig:indpotcomp} in the presence
of a relatively weak external field of 0.05 eV/\AA. The dielectric
parameters were obtained by a least square fit of the calculated
potentials to the induced electrostatic potential obtained in a
dielectric model of the film and a perfect conductor model of the
metal. Using standard electrostatics, one obtains the following
potential across the dielectric film and the perfect conductor,
\begin{equation}
\Delta\phi(z)=-2E_\mathrm{ext}\left\{\begin{array}{ll}
0, & z < z_{d}-a\\
\frac{(z-z_{d}-a)}{\epsilon_\infty}, &  z_{d}-a<z<z_{d}\\
(z-z_{d}+\frac{a}{\epsilon_\infty}), &  z>z_{d}
\end{array}\right.\label{eq:PotDielMod}
\end{equation}

This fit of the computed $\Delta\phi(z)$ using DFT-full for the NaCl
bilayer on the explicit Cu(100) surface to the model $\Delta\phi(z)$
in Eq.~\ref{eq:PotDielMod} gives $\epsilon_{\infty}=2.52$ and $a=6.26$
{\AA}. This fit using the results from DFT-PC-FF gives
$\epsilon_{\mathrm{\infty}}=2.69$ and $a=6.46$ {\AA}. In the case of a
free-standing NaCl bilayer, a fit of a similar model potential for an
free-standing dielectric film to the computed $\Delta\phi(z)$ gives
$\epsilon_{\infty}=2.5$ and $a=6.3$ {\AA}. This value of 2.5 for
$\epsilon_{\infty}$ is close to our calculated value of 2.47 for bulk
NaCl using density functional perturbation theory method in VASP.
Note that the value of 6.3 {\AA} for $a$ is much larger than the distance
of 2.87 {\AA} between the two layers of the free-standing bilayer. The
corresponding results for $N_\mathrm{l} = 3$ to 5 obtained by
DFT-FF-PC are shown in Table \ref{tab:dielmodel} and the dielectric
response of the film is well-represented by the average values of 2.64
and 1.66 {\AA} for $\epsilon_{\infty}$ and $z_{\mathrm{d}}-z_{s}$,
respectively.  Furthermore the layer dependence of $a$ is well
represented by $a(N_\mathrm{l})\approx a_{0}+N_\mathrm{l}\Delta a$ for
$N_\mathrm{l}=2$ to 5 where $a_{0}\approx$ 0.84 {\AA} and $\Delta
a\approx$ 2.81 {\AA}.

\begin{table}
\centering
\begin{tabular}{|c|c|c|c|}
\hline 
$N_\mathrm{l}$ & $\epsilon_\infty$  & $a$ ({\AA}) & $z_\mathrm{d}-z_\mathrm{s}$ ({\AA})\tabularnewline
\hline 
\hline 
2 & 2.69 & 6.46 & 1.66\tabularnewline
\hline 
3 & 2.63 & 9.21 & 1.68\tabularnewline
\hline 
4 & 2.63 & 12.12 & 1.65\tabularnewline
\hline 
5 & 2.62 & 14.89 & 1.67\tabularnewline
\hline 
\end{tabular}
\caption{Calculated dielectric model parameters for different numbers
  of NaCl layers $N_\mathrm{l}$ as obtained by a fit to the calculated
  response to an external electric field using
  DFT-FF-PC. $z_\mathrm{d}-z_\mathrm{s}$ is the distance of the
  dielectric surface from the outermost surface plane.}
\label{tab:dielmodel}
\end{table}


\begin{thebibliography}{00}
\bibitem{quasi_planar} The standard deviation for the out-of-plane
distances of the atoms in the adsorbed pentacene molecule is less than
0.05 \AA{}.
\bibitem{uncertain} The uncertainities in the experimental values are
  several tenths of an eV due to vibronic shifts~\cite{Pavetal13} and
  the estimate of the voltage drop across the NaCl film.
\end{thebibliography}
\end{document}